%% file: Lightning.tex
\documentclass[twocolumn]{aastex631}
\usepackage{amsmath}

\shorttitle{Lightning SED}
\shortauthors{Doore et al.}

\begin{document}

\title{Lightning: An X-ray to Submillimeter Galaxy SED-Fitting Code With Physically Motivated Stellar, Dust, and AGN Models}

\correspondingauthor{Keith Doore}
\email{kjdoore@uark.edu}

\author[0000-0001-5035-4016]{Keith~Doore}
\affiliation{Department of Physics, University of Arkansas, 226 Physics Building, 825 West Dickson Street, Fayetteville, AR 72701, USA}

\author[0000-0001-8473-5140]{Erik~B.~Monson}
\affiliation{Department of Physics, University of Arkansas, 226 Physics Building, 825 West Dickson Street, Fayetteville, AR 72701, USA}

\author[0000-0002-2987-1796]{Rafael~T.~Eufrasio}
\affiliation{Department of Physics, University of Arkansas, 226 Physics Building, 825 West Dickson Street, Fayetteville, AR 72701, USA}

\author[0000-0003-2192-3296]{Bret~D.~Lehmer}
\affiliation{Department of Physics, University of Arkansas, 226 Physics Building, 825 West Dickson Street, Fayetteville, AR 72701, USA}

\author[0000-0002-9202-8689]{Kristen~Garofali}
\affiliation{NASA Goddard Space Flight Center, Code 662, Greenbelt, MD 20771, USA}

\author[0000-0001-8525-4920]{Antara~Basu-Zych}
\affiliation{NASA Goddard Space Flight Center, Code 662, Greenbelt, MD 20771, USA}
\affiliation{Center for Space Science and Technology, University of Maryland Baltimore County, 1000 Hilltop Circle, Baltimore, MD 21250, USA}
\affiliation{Center for Research and Exploration in Space Science and Technology, NASA/GSFC, Greenbelt, MD 20771}



\begin{abstract}

We present an updated version of \texttt{Lightning}, a galaxy spectral energy distribution (SED) fitting code that can model X-ray to submillimeter observations. The models in \texttt{Lightning} include the options to contain contributions from stellar populations, dust attenuation and emission, and active galactic nuclei (AGN). X-ray emission, when utilized, can be modeled as originating from stellar compact binary populations with the option to include emission from AGN. We have also included a variety of algorithms to fit the models to observations and sample parameter posteriors; these include an adaptive Markov-Chain Monte-Carlo (MCMC), affine-invariant MCMC, and Levenberg-Marquardt gradient decent (\texttt{MPFIT}) algorithms. To demonstrate some of the capabilities of \texttt{Lightning}, we present several examples using a variety of observational data. These examples include (1) deriving the spatially resolved stellar properties of the nearby galaxy  M81, (2) demonstrating how X-ray emission can provide constrains on the properties of the supermassive black hole of a distant AGN, (3) exploring how to rectify the attenuation effects of inclination on the derived the star formation rate of the edge-on galaxy NGC~4631, (4) comparing the performance of \texttt{Lightning} to similar Bayesian SED fitting codes when deriving physical properties of the star-forming galaxy NGC~628, and (5) comparing the derived X-ray and UV-to-IR AGN properties from \texttt{Lightning} and \texttt{CIGALE} for a distant AGN. \texttt{Lightning} is an open-source application developed in the Interactive Data Language (IDL) and is available at \url{https://github.com/rafaeleufrasio/lightning}.

\end{abstract}

\keywords{Extragalactic astronomy (506), Galaxy properties (615), Star formation (1569), Spectral energy distribution (2129)}


\section{Introduction} \label{sec:Intro}

The light emitted from a galaxy contains a plethora of information about many physical properties of the system, ranging from its star-formation history (SFH) and dust content to the presence of an active galactic nucleus (AGN) and the properties of its supermassive black hole (SMBH). These properties are key to our current understanding of how galaxies and SMBHs formed and evolved, and, thus, the methods for deriving them from spectral energy distributions (SEDs) have been the focus of substantial work \citep[e.g.,][]{1998ApJ...509..103S, 1999A&A...350..381D, 2005ApJ...633..857D, 2008ApJS..176..438G, 2009A&A...507.1793N, 2015A&A...576A..10C, 2017ApJ...838..127I, 2018ApJ...854...62L, 2021MNRAS.505.1509S}. The overarching process of deriving the physical properties from an SED is known as SED fitting (see \citealt{2011Ap&SS.331....1W} and \citealt{2013ARA&A..51..393C} for recent reviews). At its core, this process consists of fitting a model (e.g, stellar population synthesis with dust attenuation) to the observed SED. Once a best-fit model is determined using the chosen statistical inferencing method, it can be used to infer the physical properties of the observations.

Numerous SED fitting codes currently exist today for the modeling and inferencing of galaxy properties from their SEDs. Some of the more widely cited codes include \texttt{CIGALE} \citep{2005MNRAS.360.1413B, 2019A&A...622A.103B}, \texttt{Prospector} \citep{2021ApJS..254...22J}, \texttt{MAGPHYS} \citep{2008MNRAS.388.1595D}, \texttt{BAGPIPES} \citep{2018MNRAS.480.4379C}, and \texttt{FAST} \citep{2009ApJ...700..221K}; also see \citet{2022arXiv221201915P} for a more comprehensive list. Each code was designed to help solve unanswered problems unique to their developers. Therefore, each code is unique and comes with its own set of advantages and disadvantages.

Initially, SED fitting codes were developed to use maximum-likelihood statistical inferencing methods (e.g., linear and non-linear optimization) to estimate galactic properties from optical to infrared (IR) observations (e.g., \texttt{SEDfit}, \citealt{1998AJ....115.1329S, 2012PASP..124.1208S}; \texttt{STARLIGHT}, \citealt{2005MNRAS.358..363C}; \texttt{VESPA}, \citealt{2007MNRAS.381.1252T}). These codes typically model the observations using simple stellar population (SSP) models \citep[e.g.,][]{1997A&A...326..950F, 2003MNRAS.344.1000B, 2009ApJ...699..486C, 2017PASA...34...58E} with simple parametric SFHs (e.g., exponentially declining) and attenuation. The advantage of these codes is that they are fast, simple to use, and return reliable best-fit models. However, the major drawback is that they can have difficulties computing accurate uncertainties on the physical parameters, since these parameters can be highly correlated and usually have non-Gaussian likelihoods. This difficulty is compounded as additional model components are included to account for more complex physical processes within galaxies: for example, non-parametric SFHs (see \citealt{2019ApJ...873...44C} and \citealt{2019ApJ...876....3L} for overviews on the differences between parametric and non-parametric SFHs), dust emission \citep[e.g.,][]{2007ApJ...657..810D, 2008MNRAS.388.1595D, 2012MNRAS.425.3094C, 2014ApJ...784...83D}, dusty torus emission from an AGN \citep[e.g.,][]{2006MNRAS.366..767F, 2008ApJ...685..147N, 2012MNRAS.420.2756S, 2016MNRAS.458.2288S}, and nebular emission \citep[e.g.,][]{1998PASP..110..761F, 2013RMxAA..49..137F}.

To estimate more accurate uncertainties, the next generation of SED fitting codes were developed to use a gridded Bayesian statistical inferencing method (e.g., \texttt{CIGALE}, \citealt{2005MNRAS.360.1413B, 2019A&A...622A.103B}; \texttt{FAST}, \citealt{2009ApJ...700..221K}; \texttt{MAGPHYS}, \citealt{2008MNRAS.388.1595D}). This method estimates galactic properties and their uncertainties by gridding parameter space, fitting the corresponding gridded models to the observations, and then weighting the models by their goodness of fit. The advantage of this method is that it can account for parameter degeneracies and non-Gaussian likelihoods, while still being computationally fast. However, this computational speed excludes the time to create the grid of models, which increases exponentially with the number of parameters. Therefore, sampling of the full posterior distribution becomes intractable in a reasonable amount of time for complex models with many parameters unless parameter space is sparsely sampled.

In order to better sample the parameter space of complex models, new SED-fitting codes were developed to use Bayesian sampling statistical inferencing methods which utilize Markov Chain Monte Carlo (MCMC) and/or nested sampling algorithms \citep{2004AIPC..735..395S}. This approach (e.g., \texttt{BAGPIPES}, \citealt{2018MNRAS.480.4379C}; \texttt{BayeSED}, \citealt{2012ApJ...749..123H, 2014ApJS..215....2H, 2019ApJS..240....3H}; \texttt{BEAGLE}, \citealt{2016MNRAS.462.1415C}; \texttt{Prospector}, \citealt{2021ApJS..254...22J}) has the advantage of efficiently sampling parameter space of complex models to generate posterior distributions of parameters, while taking into account any prior information on the parameters. Additionally, models can be changed without any computational cost unlike the gridded Bayesian methods, which require the entire grid of models to be recomputed. However, the disadvantage of the Bayesian sampling approach is that sampling the posterior distribution can take significantly longer computational times on a per-SED basis.

Some next generation SED fitting codes are trying to bridge the gap between parameter estimation and computational speed using machine learning. For these codes (e.g., \citealt{2019MNRAS.490.5503L}; \texttt{mirkwood}, \citealt{2021ApJ...916...43G}), machine-learning models are trained to learn the relationship between input observations and inferred properties using synthetic galaxy SEDs generated by cosmological simulations. The major advantage of these codes is that, once trained, fitting a new input SED is incredibly fast and derived parameters can be more accurate than the fully Bayesian approach \citep{2021ApJ...916...43G}. However, in their current state of development, machine-learning SED-fitting codes come with a few serious drawbacks. The first is that they have an over-reliance on theoretical models to explain how real galaxies should appear. Since they typically utilize a variety of cosmological simulations, it can be difficult to create a complete training set that is truly representative of all observed galaxies \citep{2014MNRAS.445..175G, 2015MNRAS.446..521S, 2015ARA&A..53...51S}. Additionally, overtraining (which can lead to overfitting) can occur when an appropriate test set or cross-validation set is not utilized. This can lead to galaxy property estimates that have high precision, which is a direct result of overfitting rather than a correct uncertainty estimate. Finally, these codes cannot handle missing observations that are commonly present in typical SEDs without retraining. This comes at a significant computational cost if the sample that is to be fit has a variety of observations.

Motivated to have a computationally fast yet fully Bayesian code, we developed the SED-fitting code \texttt{Lightning}. Originally developed to derive the SFHs needed to empirically calibrate the X-ray luminosity function (XLF) of X-ray binary (XRB) populations \citep{2017ApJ...851...10E, 2017ApJ...851...11L, 2020ApJS..248...31L, 2022ApJ...930..135L, 2022ApJ...926...28G}, \texttt{Lightning} has since been utilized to check for enhanced star formation and AGN activity in protoclusters \citep{2021ApJ...919...51M}, model local analogs to high-redshift galaxies \citep{2021ApJ...921..130M}, investigate the inclination-dependence of derived SFHs \citep{2021ApJ...923...26D, 2022ApJ...931...53D}, and provide evidence in favor of density wave theory \citep{2022MNRAS.512..366A}. Written in IDL, the newest updates to \texttt{Lightning} now allow for modeling of photometric SEDs from the X-rays to submillimeter using efficient MCMC algorithms to fit physical models that account for any combination of stellar, dust, and AGN emission. In Section~\ref{sec:Models}, we describe these physical models and their dependencies. The statistical inferencing methods that fit these models to input SEDs are described in Section~\ref{sec:Inference}. In Section~\ref{sec:Examples}, we demonstrate the capability of \texttt{Lightning} by applying it to a variety of examples. Finally, we summarize and discuss future planned additions for \texttt{Lightning} in Section~\ref{sec:Summary}.

\texttt{Lightning} is an open-source, well-documented, and publicly available SED fitting code available at \url{https://github.com/rafaeleufrasio/lightning}. Since \texttt{Lightning} has been in development over the past several years, we include the references to past works that first described each feature (i.e., models and statistical inferencing methods) and the motivation for their implementation in Table~\ref{tab:features}. In Sections~\ref{sec:Models} and~\ref{sec:Inference}, we reiterate the details from these past works and clarify all assumptions for replicability.
\input{table_feature.tex}

\section{Physical Models} \label{sec:Models}

In this section, we describe the variety of physical models available in \texttt{Lightning} to account for any combination of stellar, dust, and AGN emission. To help clarify the free parameters corresponding to each model, we give a description of the parameters, their units, and their allowed range in Table~\ref{tab:SEDParams}.

\input{table_SEDparams.tex}

\subsection{Stellar Emission Models}

\subsubsection{Simple Stellar Populations}

\begin{figure}[t!]
\centering
\includegraphics[width=8.5cm]{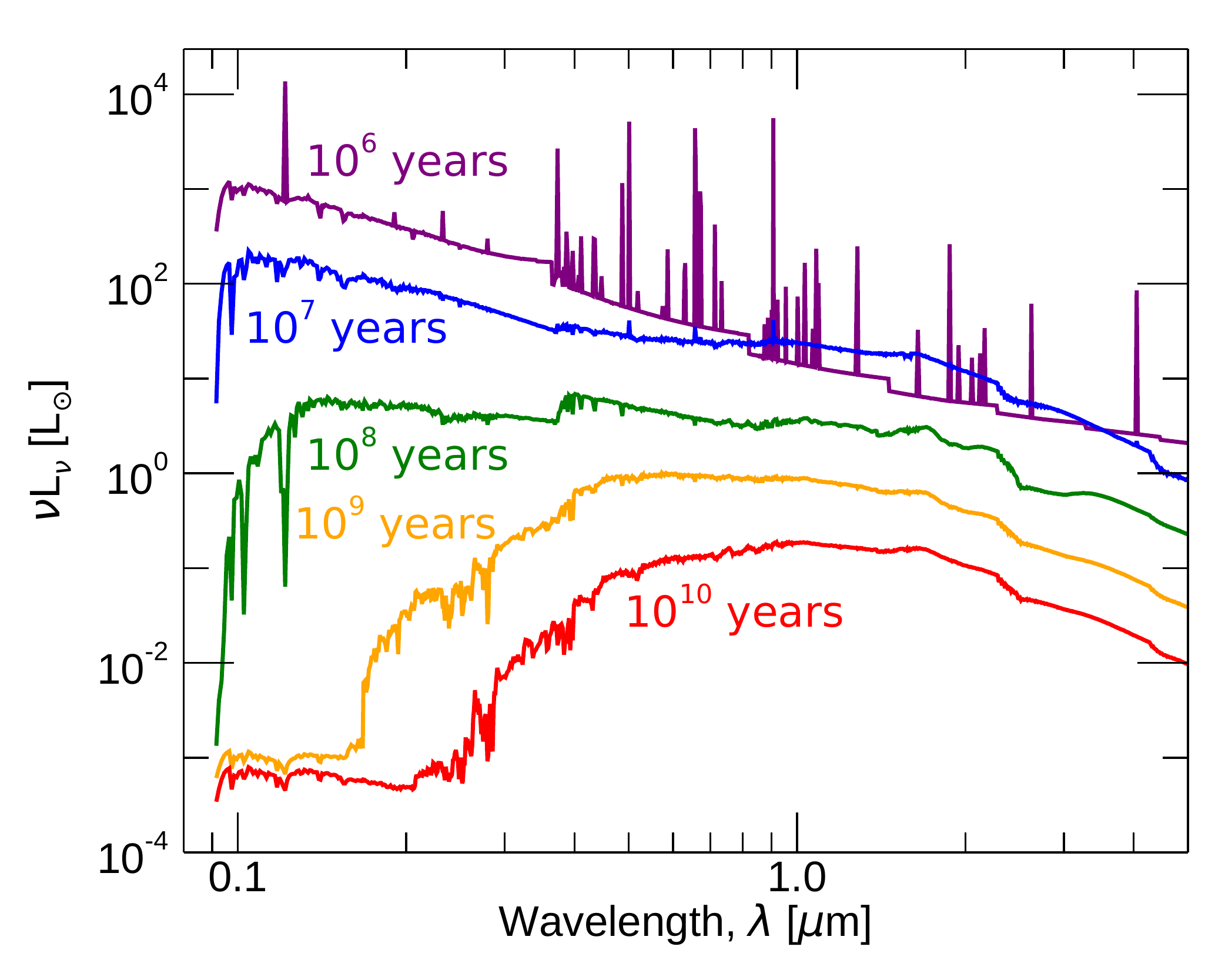}
\caption{
The \texttt{P\'EGASE} SSP SEDs used in \texttt{Lightning} at various ages after ZAMS for a metallicity of $Z = 0.02$ and initial mass of 1~$M_\odot$. For an age of 1~Myr, nebular emission lines can be clearly seen, while the older displayed ages do not have any lines due to our simplifying assumption. Additionally, as the population ages, the overall bolometric luminosity decreases, and the peak wavelength of the emission shifts from the UV into the NIR.
}
\label{fig:ssp}
\end{figure}

The intrinsic UV-to-IR stellar emission in \texttt{Lightning} is generated using the SSPs from the population synthesis code \texttt{P\'EGASE} \citep{1997A&A...326..950F}. The SSPs, which are generated assuming the \cite{2001MNRAS.322..231K} initial mass function (IMF), are instantaneous bursts of star formation normalized to a unit star formation rate (SFR) of 1 $M_\odot\ {\rm yr}^{-1}$. We allow for a wide range of metallicity options when generating the stellar populations (0.001, 0.004, 0.008, 0.01, 0.02, 0.05, and 0.1 in terms of~$Z$) along with the option to include the nebular extinction and emission from \texttt{P\'EGASE} (see Section 2.4 of \citealt{1997A&A...326..950F}). While the nebular extinction affects SEDs of all ages, nebular emission is only added to stellar populations with ages $<$50 Myr. We make this simplifying assumption since there is minimal ionizing flux, which causes the nebular emission, for populations with ages $>$50 Myr \citep{2002MNRAS.337.1309S, 2017ApJ...840...44B}. In Figure~\ref{fig:ssp}, we show some example SSPs used in \texttt{Lightning} at different ages for a metallicity of $Z = 0.02$ (i.e., solar metallicity).

\subsubsection{Star Formation History} \label{sec:SFH}

To model complex SFHs while remaining computationally fast, \texttt{Lightning} assumes a simple non-parametric SFH. Continuing with the original description in \citet{2017ApJ...851...10E}, we define this to be a piece-wise constant SFH, where the free parameters for the SFH are the SFRs ($\psi_j$) within the user-defined age bins. This is given in analytical form as a function of stellar age $t$ by
\begin{equation}
    \psi(t) = \psi_j \qquad {\rm for} \qquad t_j < t < t_{j+1},
\end{equation}
where $t_j$ and $t_{j+1}$ are the respective lower and upper age bin boundaries of the $j$th bin. The advantage of normalizing by SFR, versus stellar mass like other SED fitting codes, is that any bias toward rising SFHs is prevented, while still allowing for bursty SFHs \citep{2019ApJ...876....3L}.

To compute the intrinsic, rest-frame composite stellar spectrum for the $j$th age bin, $\tilde{L}_{\nu, j}^{\star}(\nu)$\footnote{When symbolizing intrinsic emission (i.e., no attenuation), we include a tilde over the variable to clarify that it is intrinsic. Emission variables without a tilde signify attenuation/absorption has been applied.}, the SSPs are integrated over the age bin after interpolating the SSP evolution to a common time grid using a user-defined time step (e.g., 0.5 Myr). The total composite stellar spectrum for all ages, $\tilde{L}_\nu^{\star}(\nu)$, is then given by
\begin{equation}
    \tilde{L}_\nu^{\star}(\nu) = \sum_{j=1}^{n} \psi_j \tilde{L}_{\nu, j}^{\star}(\nu),
\end{equation}
where $\tilde{L}_{\nu, j}^{\star}(\nu)$ is by construction normalized per unit SFR, specifically 1 $M_\odot$~yr$^{-1}$.

\begin{figure}[t!]
\centering
\includegraphics[width=8.5cm]{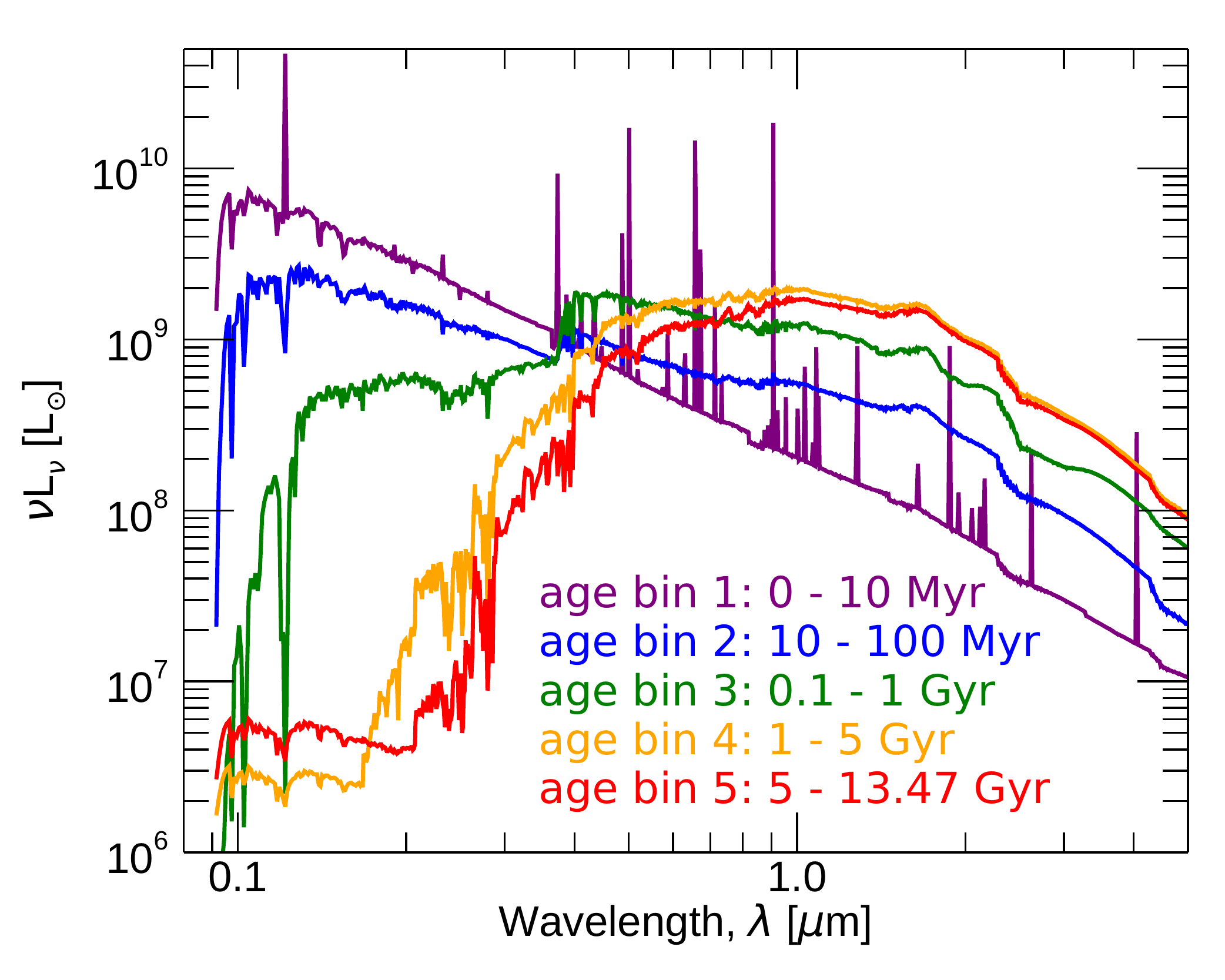}
\caption{
Example composite stellar spectrum for a metallicity of $Z = 0.02$ using the default set of age bins in \texttt{Lightning}. Similarly to the SSPs in Figure~\ref{fig:ssp}, the nebular emission lines can be clearly seen in the youngest bin, while the older bins lack any obvious emission features.
}
\label{fig:AgeBins}
\end{figure}

In Figure~\ref{fig:AgeBins}, we show composite stellar spectra for a metallicity of $Z = 0.02$ using the default set of age bins in \texttt{Lightning} (0--10 Myr, 10--100 Myr, 0.1--1 Gyr, 1--5 Gyr, 5--13.6 Gyr). These age bins were chosen such that the youngest age bin encapsulates the stellar population able to emit the majority of the ionizing flux, while the second bin includes the stellar population which generates the remaining bulk of the UV emission. Finally, the last three age bins were selected to have similar bolometric luminosities as the second bin in the case of a constant SFH.

\subsubsection{X-ray Binary Model}

We include stellar X-ray emission from compact object binaries in \texttt{Lightning} with a power-law spectral model given by
\begin{equation} \label{eq:PowerLaw}
    \tilde{L}_\nu \propto \exp\left(h \nu / E_{\rm cut}\right) (h \nu)^{1-\Gamma},
\end{equation}
where we assume a photon index of $\Gamma = 1.8$ and cutoff energy of $E_{\rm cut} = 100$ keV. We set the normalization of the power-law by its rest-frame 2--10 keV luminosity $\tilde{L}_{\rm X}$, which we model to include contributions from both high-mass XRB (HMXB) emission and low-mass XRB (LMXB) emission:
\begin{equation} \label{eq:LxXRB}
	\tilde{L}_{\rm X} = \tilde{L}^{\rm HMXB}_{\rm X} + \tilde{L}^{\rm LMXB}_{\rm X}.
\end{equation}
Our models of $\tilde{L}^{\rm HMXB}_{\rm X}$ and $\tilde{L}^{\rm LMXB}_{\rm X}$ are calculated using the empirical parameterizations with stellar age from \citet{2022ApJ...926...28G}:
\begin{equation} \label{eq:hmxb}
	\frac{\tilde{L}^{\rm HMXB}_{\rm X}}{M_{\star}}(t) = -0.24(\log_{10}(t) - 5.23)^2 + 32.54,
\end{equation}
and
\begin{equation} \label{eq:lmxb}
	\frac{\tilde{L}^{\rm LMXB}_{\rm X}}{M_{\star}}(t) = -1.21(\log_{10}(t) - 9.32)^2 + 29.09,
\end{equation}
where $\tilde{L}_{\rm X}/M_\star$ has units of ${\rm erg\ s^{-1}}\ M_\odot^{-1}$ and $t$ is the stellar age in yr. While studies have shown that the luminosity of HMXBs depends on metallicity \citep[e.g.,][]{2021ApJ...907...17L}, we do not currently implement any metallicity dependence in our X-ray binary model. The age-dependent relationship from \citet{2022ApJ...926...28G} was derived for galaxies with metallicities ranging from 0.40--1.16 $Z_{\odot}$. For larger metallicities, $\tilde{L}_X/M_\star$ may thus be slightly overestimated for $t \lesssim 100$ Myr, while for lower metallicites it may be similarly underestimated.

To calculate the scaling parameter $\tilde{L}_{\rm X}$, we first derive the HMXB and LMXB contribution from each age bin utilizing Equations~\ref{eq:hmxb} and~\ref{eq:lmxb}. We calculate the stellar mass of each bin ($M_{\star, j}$) as
\begin{equation}
    M_{\star, j} = \psi_j M_{\star, j}^{\rm coeff},
\end{equation}
where $M_{\star, j}^{\rm coeff}$ is the coefficient that converts SFR into stellar mass. We then calculate $\tilde{L}_{\rm X}/M_\star$ at the mean stellar age of each SFH bin, multiply by the stellar mass in the bin, and sum the contributions from each bin to derive the total contributions from the HMXBs and LMXBs, which are then incorporated into Equation~\ref{eq:LxXRB} to determine $\tilde{L}_{\rm X}$ and finally the X-ray luminosity spectrum.

\subsection{AGN Emission Models}

\subsubsection{AGN UV-to-IR Models} \label{sec:skirtor}

\begin{figure}[t!]
\centering
\includegraphics[width=8.5cm]{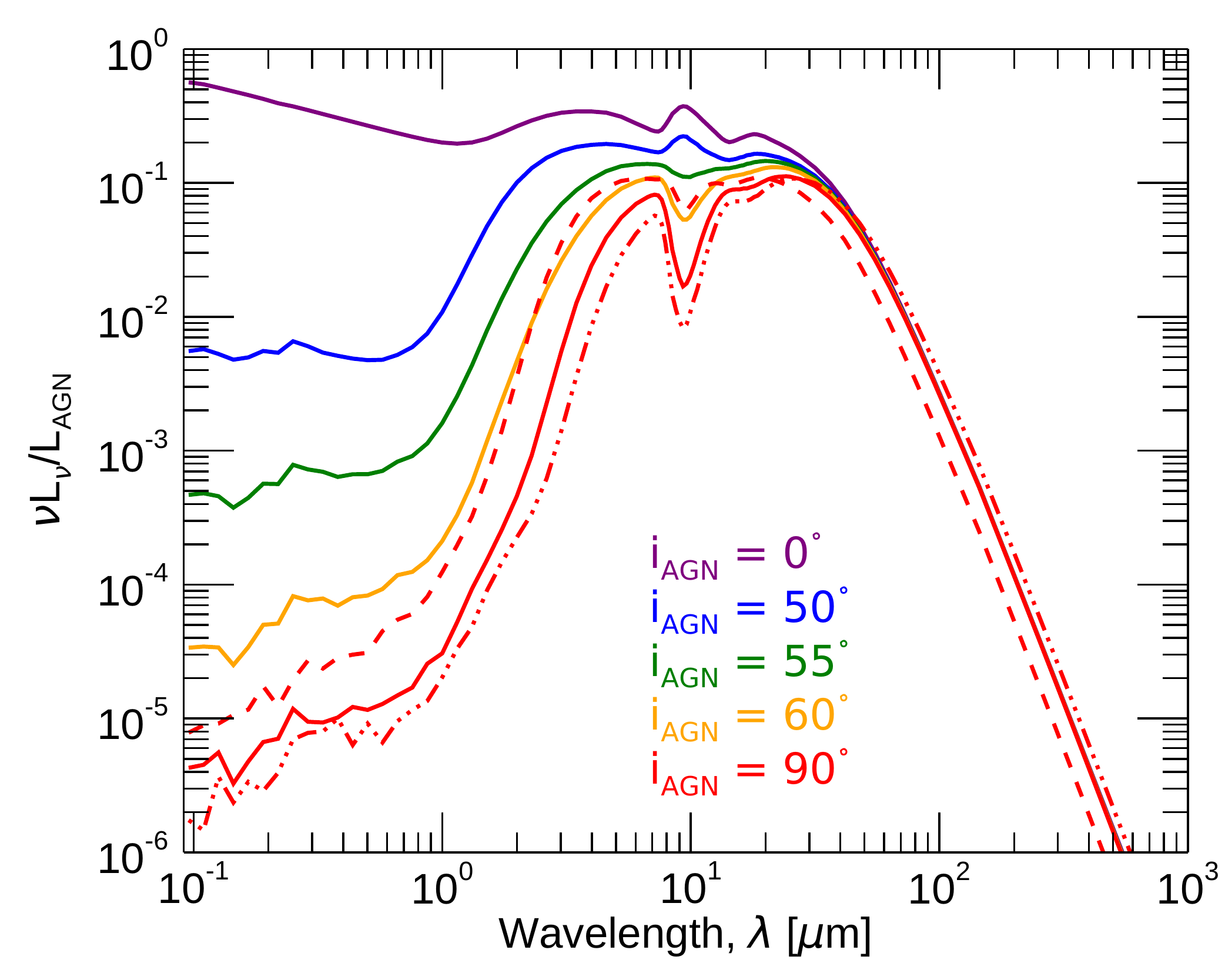}
\caption{
Examples of the SKIRTOR AGN emission model generated by \texttt{Lightning} for a range of inclinations with $\tau_{9.7} = 7$ normalized by $L_{\rm AGN}$. The dashed and dashed-dotted lines for the $i_{\rm AGN} = 90^\circ$ example are models where $\tau_{9.7} = 3$ and $\tau_{9.7} = 11$, respectively. We only show the variation in $\tau_{9.7}$ for the edge-on example, since $\tau_{9.7}$ is the edge-on optical depth of the torus. Therefore, changing its value for face-on to moderately inclined viewing angles minimally effects the model. 
}
\label{fig:UVIRAGN}
\end{figure}

\texttt{Lightning} uses the SKIRTOR library of UV-to-IR AGN SED templates \citep{2012MNRAS.420.2756S,2016MNRAS.458.2288S}, which consist of a broken power law accretion disk component and a clumpy two-phase dusty torus that reprocesses light from the accretion disk into the NIR. Our implementation uses the default accretion disk power law from the SKIRTOR library, where
\begin{equation}
\lambda \tilde{L}_\lambda \propto
	\begin{cases}
		\lambda^{1.2} & 0.001\ \micron \leq \lambda \leq 0.01\ \micron \\
		\lambda^{0} & 0.01\ \micron < \lambda \leq 0.1\ \micron \\
		\lambda^{-0.5} & 0.1\ \micron < \lambda \leq 5\ \micron \\
		\lambda^{-3} & 5\ \micron < \lambda \leq 50 \micron \\
	\end{cases}.
\end{equation}
The full SKIRTOR templates have six free parameters: $\tau_{9.7}$, the edge-on optical depth of the torus at $9.7\ \micron$; $p$, the power law index for the radial dust density gradient; $q$, the power law index for the polar dust density gradient; $\Delta$, the opening angle of the dusty cone of the torus; $R$, the ratio of the torus' inner and outer radii; and $i_{\rm AGN}$, the inclination angle from the pole to the line of sight. To simplify the SKIRTOR models and allow us to sample the parameter space and interpolate between models, our implementation in \texttt{Lightning} only allows for $\tau_{9.7}$ and $\cos i_{\rm AGN}$ to be free parameters. We fix $p=1$ and $q=0$ (i.e., there is no polar dependence of the dust density) as in \citet{2016MNRAS.458.2288S}, and fix $\Delta=40^{\circ}$ based on their findings of typical covering factors in the range of $0.6-0.7$. At the moderate covering factor of $\sin40^{\circ} \approx 0.64$ that we assume, $R$ has only a small impact on the luminosity of the torus, so we fix $R=20$. To implement these simplified models, we linearly interpolate the original gridded models both in $\tau_{9.7}$ and $\cos i_{\rm AGN}$-space for the desired free parameter value. Then if no X-ray AGN model is used, the UV-to-IR AGN model is scaled using its anisotropic integrated luminosity, $L_{\rm AGN}$, as another free parameter. Examples of these UV-to-IR AGN emission models for different inclinations are shown in Figure~\ref{fig:UVIRAGN}.

We note that the UV-optical light that escapes the torus is subject to being attenuated by the host galaxy dust (see Section~\ref{sec:DustAtten}). When energy balance is enabled (as discussed in Section~\ref{sec:DustEmission}), we integrate this attenuated light over all lines of sight\footnote{For simplicity, \texttt{Lightning} currently does not support the simultaneous usage of the SKIRTOR UV-to-IR AGN model and the inclination-dependent attenuation model, due to the fact that the lines of sight of the two models are not required to align (i.e., $\cos i \ne \cos i_{\rm AGN}$), which complicates the energy balance calculation.} and add it to the bolometric luminosity of the attenuated stellar light used to scale the dust model. However, fully energetic self-consistency is not maintained by the AGN model. The ISM is assumed to be opaque to the ionizing Lyman-continuum radiation from the AGN, and the ionizing flux from the AGN does not currently contribute to the nebular emission component, which is built into our \texttt{P\'EGASE} models.

\subsubsection{AGN X-ray Models} \label{sec:XrayAGN}

X-ray observations can place powerful constraints on the bolometric luminosity of AGN, and as such they are very useful in AGN SED modeling. In \texttt{Lightning}, we provide two different models to generate the X-ray emission from the AGN component. 

Since X-ray spectra of AGN are often empirically described by power-law models, we provide this as a basic option in \texttt{Lightning}. The equation for the power-law is given in Equation~\ref{eq:PowerLaw}, where we use a fixed photon index of $\Gamma = 1.8$ and cutoff energy of $E_{\rm cut} = 300~{\rm keV}$. To connect the X-ray model to the UV-optical AGN component, we use the \citet{2017A&A...602A..79L} $\tilde{L}_{2\rm\ keV}-\tilde{L}_{2500}$ relationship, which is an empirically calibrated relationship between the intrinsic monochromatic luminosities of luminous AGN at $2\rm~keV$ and $2500\rm~\AA$. Therefore, the power-law X-ray and UV-to-IR AGN models are both scaled using $L_{\rm AGN}$ as the only free parameter. We note that while \citet{2017A&A...602A..79L} show that their relationship has a dispersion of $\sim$0.21~dex, we do not currently provide any increased flexibility by allowing for a deviation from the best-fit $\tilde{L}_{2\rm\ keV}-\tilde{L}_{2500}$ relation. Instead, one of our general philosophies in \texttt{Lightning} is to provide model flexibility with physical parameters wherever possible.
\begin{figure*}[t!]
\centering
\includegraphics[width=18cm]{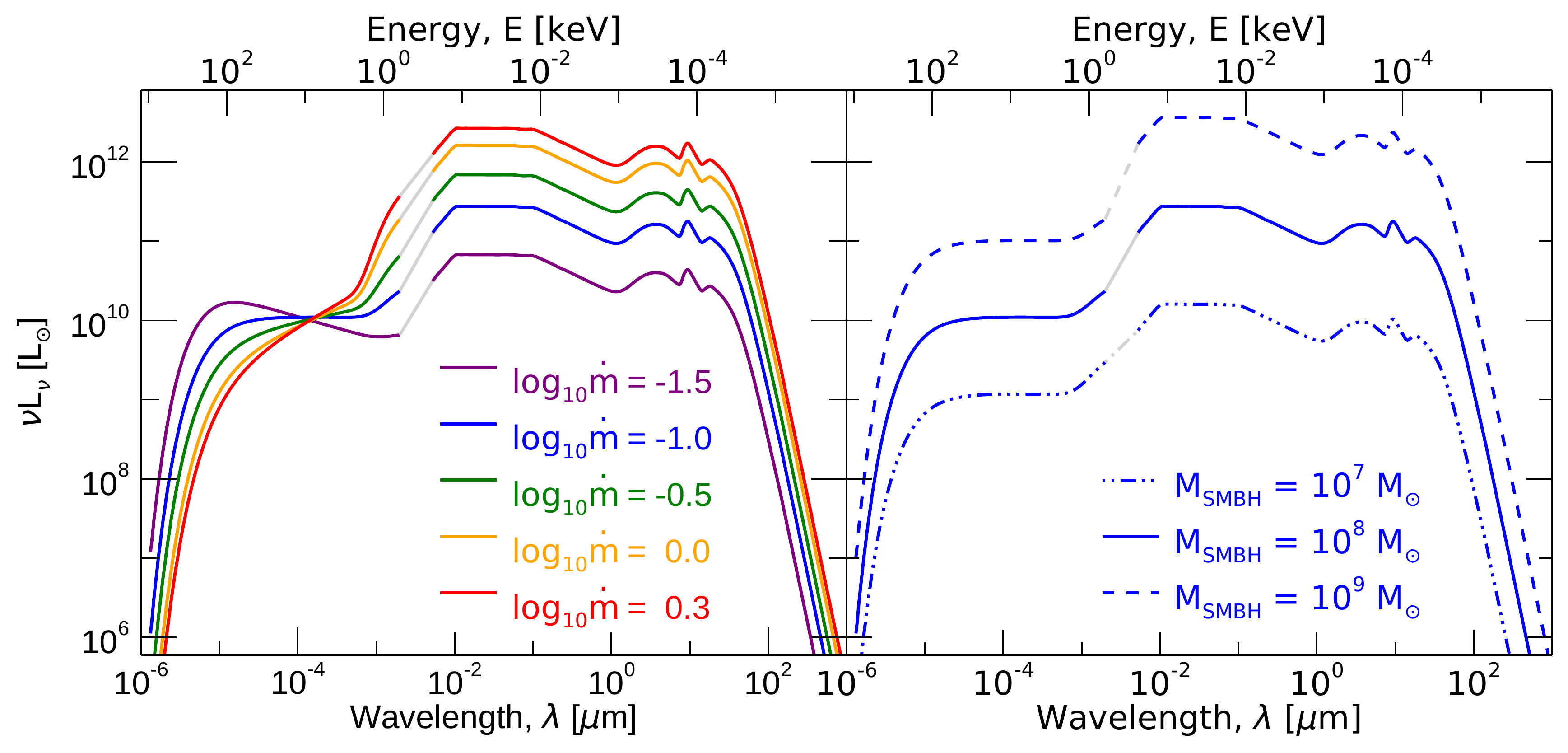}
\caption{
Examples of the \texttt{qsosed} AGN X-ray emission model connected with the SKIRTOR UV-to-IR model generated by \texttt{Lightning}. The SKIRTOR model shape is fixed using $i_{\rm AGN} = 0^\circ$ and $\tau_{9.7} = 7$ with the normalization being set by the \texttt{qsosed} model. The variation in color in the left panel corresponds to a change in $\log_{10} \dot m$ with fixed $M_{\rm SMBH} = 10^8\ M_\odot$, while different line styles in the right panel correspond to variations in $M_{\rm SMBH}$ with fixed $\log_{10} \dot m = -1$. The light gray line segments between $\lambda = 2$--5~nm ($E = 0.24$--0.62 keV) show the linear connection of the two models for visualization purposes at the edge of the X-rays. From these examples, it can be seen that $\log_{10} \dot m$ affects the shape of the X-ray emission, while $M_{\rm SMBH}$ mainly affects the normalization.  
}
\label{fig:XrayAGN}
\end{figure*}

To account for the scatter in the $\tilde{L}_{2\rm\ keV}-\tilde{L}_{2500}$ relationship in a physically-motivated way, we provide an implementation of the \texttt{qsosed} X-ray AGN models from \citet{2018MNRAS.480.1247K}, which reproduces the \citet{2017A&A...602A..79L} relationship, including its scatter, as a function of SMBH mass and Eddington ratio. These models include an accretion disk component and two Comptonizing regions, which produce the X-ray spectrum. Since these models include optical emission from the accretion disk, the relationship between $\tilde{L}_{2500}$ and $\tilde{L}_{2\rm\ keV}$ is encoded in two model parameters: $M_{\rm SMBH}$, the SMBH mass, and $\dot m$, the Eddington ratio of the AGN. \citet{2018MNRAS.480.1247K} show that their models reproduce the \citet{2017A&A...602A..79L} relationship, explaining the dispersion around the relationship due to variance of $M_{\rm SMBH}$ and $\dot m$ among AGNs. Thus, when this model is used, we normalize the UV-to-IR AGN model (see Section~\ref{sec:skirtor}) to the same $\tilde{L}_{2500}$ as the X-ray model. This allows the free parameters of $M_{\rm SMBH}$ and $\dot m$ to set the normalization of the entire X-ray-to-IR AGN model and encapsulate the variation in the $\tilde{L}_{2\rm\ keV}-\tilde{L}_{2500}$ relationship, thereby replacing $L_{\rm AGN}$ as a free parameter. Due to its physically-motivated nature, we set the \texttt{qsosed} model to the default X-ray AGN model in \texttt{Lightning} and recommend its usage. However, we note that this model is most appropriate for luminous, high-accretion rate systems ($\log_{10} \dot m$ ranges from $-1.5$ to $0.3$) and is not appropriate for low-luminosity AGN and Compton-thick AGN, the latter of which require more careful and complicated X-ray modeling with reflection components.

We show some examples of the connected \texttt{qsosed} and SKIRTOR UV-to-IR AGN models in Figure~\ref{fig:XrayAGN}. While the \texttt{qsosed} model extends to optical wavelengths and is used to normalize the UV-to-IR SKIRTOR model, we limit it to $\lambda < 2$~nm rather than directly joining the two models across the extreme-UV wavelength range. Therefore, the light gray line segments in Figure~\ref{fig:XrayAGN} show a linear interpolation between the two models for visualization purposes. 

This implementation of the connection between the X-ray, optical, and IR AGN emission is a half-step to a fully energetically self-consistent model, in which the X-ray and IR AGN spectra are generated from the same assumed torus model and accretion disk spectrum. For steps toward connecting X-ray and IR spectral models using the same torus, see e.g. \citet{2020ApJ...897....2T}.

\subsection{Dust Attenuation Models} \label{sec:DustAtten}

To account for the variety of attenuation laws between and within galaxies, we include several prescriptions for attenuation in \texttt{Lightning}. For the UV-to-IR, these include the original and a variable form of the \citet{2000ApJ...533..682C} attenuation curve and the inclination-dependent attenuation curve from \citet{2021ApJ...923...26D}. For the X-rays, these include the \texttt{tbabs} absorption model \citep{2000ApJ...542..914W} and the \texttt{Sherpa atten} model from \citet{1994AJ....107.2108R}.\footnote{Absorption is the dominant contribution to the attenuation of high energy photons such as X-rays. Therefore, it is conventional to only model absorption rather than attenuation, which includes both absorption and scatter.}

\subsubsection{\texorpdfstring{\cite{2000ApJ...533..682C}}{Calzetti et al. (2000)} Attenuation}

\begin{figure*}[t!]
\centering
\includegraphics[width=17cm]{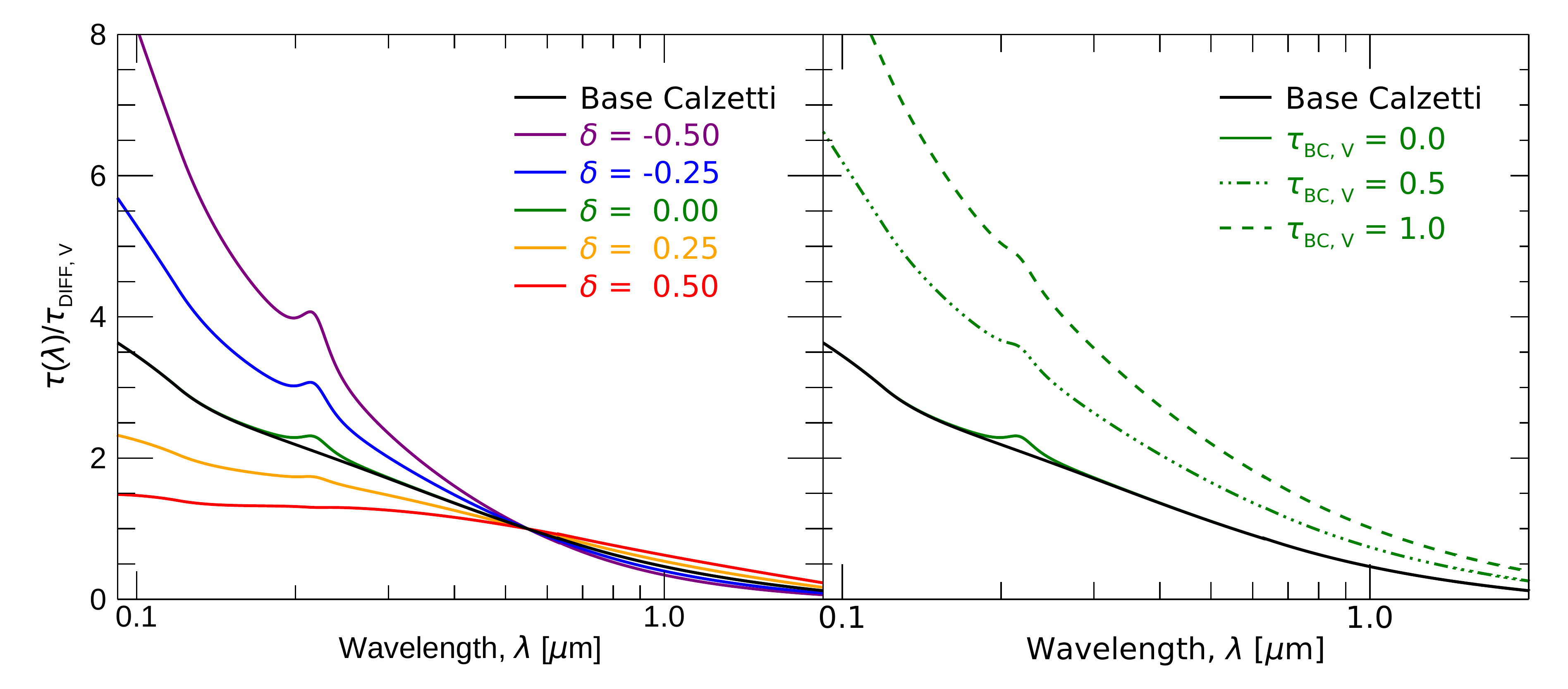}
\caption{
Example modified \citet{2000ApJ...533..682C} attenuation curves generated by \texttt{Lightning}, normalized by $\tau_{{\rm DIFF},V}$. The black curve gives the base \citet{2000ApJ...533..682C} attenuation curve (i.e., $\delta = 0$, $\tau_{{\rm BC},V} = 0$, and no UV bump feature). The variation in color in the left panel corresponds to a change in $\delta$ with fixed $\tau_{{\rm BC},V} = 0$, while different line styles in the right panel correspond to variations in $\tau_{{\rm BC},V}$ with fixed $\delta = 0$.
}
\label{fig:calzetti}
\end{figure*}

We implement the commonly used \cite{2000ApJ...533..682C} attenuation curve as the default in \texttt{Lightning}. The general details and format of this attenuation curve as implemented in \texttt{Lightning} are presented in Section~3.2 of \citet{2017ApJ...851...10E}. To briefly summarize these details, we used the \cite{2000ApJ...533..682C} curve as linearly extrapolated by \citet{2009A&A...507.1793N} at 1200~\AA\ to extend to the Lyman limit (912~\AA). To allow for more flexibility, we include the optional variable slope and 2175~\AA\ bump feature modifications as presented in \citet{2009A&A...507.1793N}. Reformatting to use the optical depth, rather than attenuation in magnitudes, this variable diffuse dust attenuation curve is defined as
\begin{equation}
    \tau_{\rm DIFF}(\lambda) = \frac{\tau_{{\rm DIFF},V}}{4.05} \Big(k^\prime(\lambda) + D(\lambda)\Big) \Bigg(\frac{\lambda}{0.55\ \mu {\rm m}}\Bigg)^\delta,
\end{equation}
where $\tau_{\rm DIFF}$ is the optical depth of the diffuse dust at wavelength $\lambda$, $\tau_{{\rm DIFF},V}$ is the $V$-band (0.55~$\mu$m) normalization, $k^\prime(\lambda)$ is the \cite{2000ApJ...533..682C} attenuation curve, $D(\lambda)$ is the functional Drude profile parameterizing the 2175~\AA\ bump feature, and $\delta$ is the parameter controlling the variable slope. The Drude profile is defined as
\begin{equation}
    D(\lambda) = \frac{E_b (\lambda\ \Delta \lambda)^2}{(\lambda^2 - \lambda^2_0)^2 + (\lambda\ \Delta \lambda)^2},
\end{equation}
where we assume a bump strength of $E_b = 0.85 - 1.9 \delta$ following the results of \citet{2013ApJ...775L..16K}, a bump FWHM of $\Delta \lambda = 350~{\rm \AA}$, and a central bump wavelength of $\lambda_0 = 2175~{\rm \AA}$. Finally, we also include an optional birth-cloud attenuation component given by
\begin{equation}
    \tau_{\rm BC}(\lambda) = \tau_{{\rm BC}, V} \Bigg( \frac{\lambda}{0.55\ \mu {\rm m}} \Bigg)^{-1},
\end{equation}
which is only applied to the youngest defined SFH age bin ($\psi_1$).\footnote{We only recommend using birth cloud attenuation when $\psi_1$ has an upper age bin boundary of 10 Myr or less, as stars older than 10 Myr typically have migrated out of or cleared their birth cloud.} Combining the diffuse dust and birth-cloud components, the effective optical depth of the attenuation curve for a given age bin $j$ is given by
\begin{equation}
    \tau_j(\lambda) = \tau_{\rm DIFF}(\lambda) + \delta_{j1} \tau_{\rm BC}(\lambda),
\end{equation}
where $\delta_{j1}$ is the Kronecker delta (not to be confused with the variable slope parameter $\delta$). Therefore, the variable \citet{2000ApJ...533..682C} attenuation has up to three free parameters ($\tau_{{\rm DIFF},V}$, $\delta$, and $\tau_{{\rm BC},V}$) that define the shape of the curve.

In Figure~\ref{fig:calzetti}, we show several modified \citet{2000ApJ...533..682C} attenuation curves as colored lines to compare with the base \citet{2000ApJ...533..682C} attenuation curve (i.e., $\delta = 0$, $\tau_{{\rm BC},V} = 0$, and no UV bump feature), which is shown in black. The variation in color corresponds to a change in $\delta$ (left panel), while different line styles correspond to variations in $\tau_{{\rm BC},V}$ (right panel). This comparison clearly shows how the variable slope parameter $\delta$ affects the slope of UV attenuation. Additionally, the inclusion of the birth cloud attenuation increases the amount of attenuation at all wavelengths.

\subsubsection{Inclination-dependent Attenuation}

To allow for more accurate attenuation in disk galaxies, an inclination-dependent attenuation curve is also included in \texttt{Lightning}, the description of which is presented in detail in Section~4.3 of \citet{2021ApJ...923...26D}. To give a brief description, the prescription is based on the \citet{2004A&A...419..821T} inclination-dependent attenuation curves, as updated by \citet{2011A&A...527A.109P}, which assume that disk galaxies are comprised of three components, a young thin disk, an old thick disk, and an old dustless bulge. The equation defining the curves in \citet{2004A&A...419..821T} was restructured to depend on the intrinsic properties of these three components rather than their observable properties. This resulted in the attenuation curve being defined by
\begin{multline}
\Delta m_{\lambda} = -2.5 \log \Bigg(\frac{r^{0,\rm old}}{1  + B/D}\ 10^\frac{\Delta m_{\lambda}^{\rm disk}(i,\tau_B^f)}{-2.5} \\
+ (1-r^{0,\rm old})(1-F f_{\lambda}) 10^\frac{\Delta m_{\lambda}^{\rm tdisk}(i,\tau_B^f)}{-2.5} \\
+ \Big(r^{0,\rm old} - \frac{r^{0,\rm old}}{1  + B/D}\Big) 10^\frac{\Delta m_{\lambda}^{\rm bulge}(i,\tau_B^f)}{-2.5}  \Bigg).
\end{multline}
Here, $\Delta m_{\lambda}$ is the attenuation in magnitudes at a given wavelength $\lambda$. $r^{0,\rm old}$ is the fraction of intrinsic flux density from the old stellar components (i.e., thick disk and bulge) compared to the total intrinsic flux density. $B/D$ is the intrinsic bulge-to-thick disk ratio. $\Delta m_{\lambda}^{\rm disk}(i,\tau_B^f)$, $\Delta m_{\lambda}^{\rm tdisk}(i,\tau_B^f)$, and $\Delta m_{\lambda}^{\rm bulge}(i,\tau_B^f)$ are the attenuations from the diffuse dust, parameterized by fifth-order polynomials in \citet{2011A&A...527A.109P}, which are functions of inclination, $i$, for tabulated values of the $B$-band face-on optical depth as seen through the center of the galaxy, $\tau_B^f$, and wavelength for the disk, thin disk, and bulge, respectively. Finally, $F$ is the birth cloud clumpiness factor of the thin disk, and $f_{\lambda}$ is a tabulated function of wavelength that provides $F$ its wavelength dependence.

The restructuring of the original \citet{2004A&A...419..821T} equation to intrinsic properties was intentional so that the non-parametric SFH in \texttt{Lightning} could be used to effectively eliminate $r^{0,\rm old}$ as a free parameter. This was done by making $r^{0,\rm old}$ a binary parameter, where each SFH age bin is given its own value for $r^{0,\rm old}$ based on its age. A value of 0 indicates that the given SFH bin is to be considered part of the young stellar population (e.g., $t < 500$ Myr), while a value of 1 considers it part of the old stellar population (e.g., $t > 500$ Myr). Therefore, with $r^{0,\rm old}$ tied to the SFH ages, the other four parameters ($i$, $\tau_B^f$, $B/D$, and $F$) are the free parameters defining the shape of the inclination-dependent attenuation curve. Examples of these inclination-dependent attenuation curves can be found in Figures~7 and~8 of \citet{2021ApJ...923...26D}.

\subsubsection{X-ray Absorption}

\begin{figure*}[t!]
\centering
\includegraphics[width=17cm]{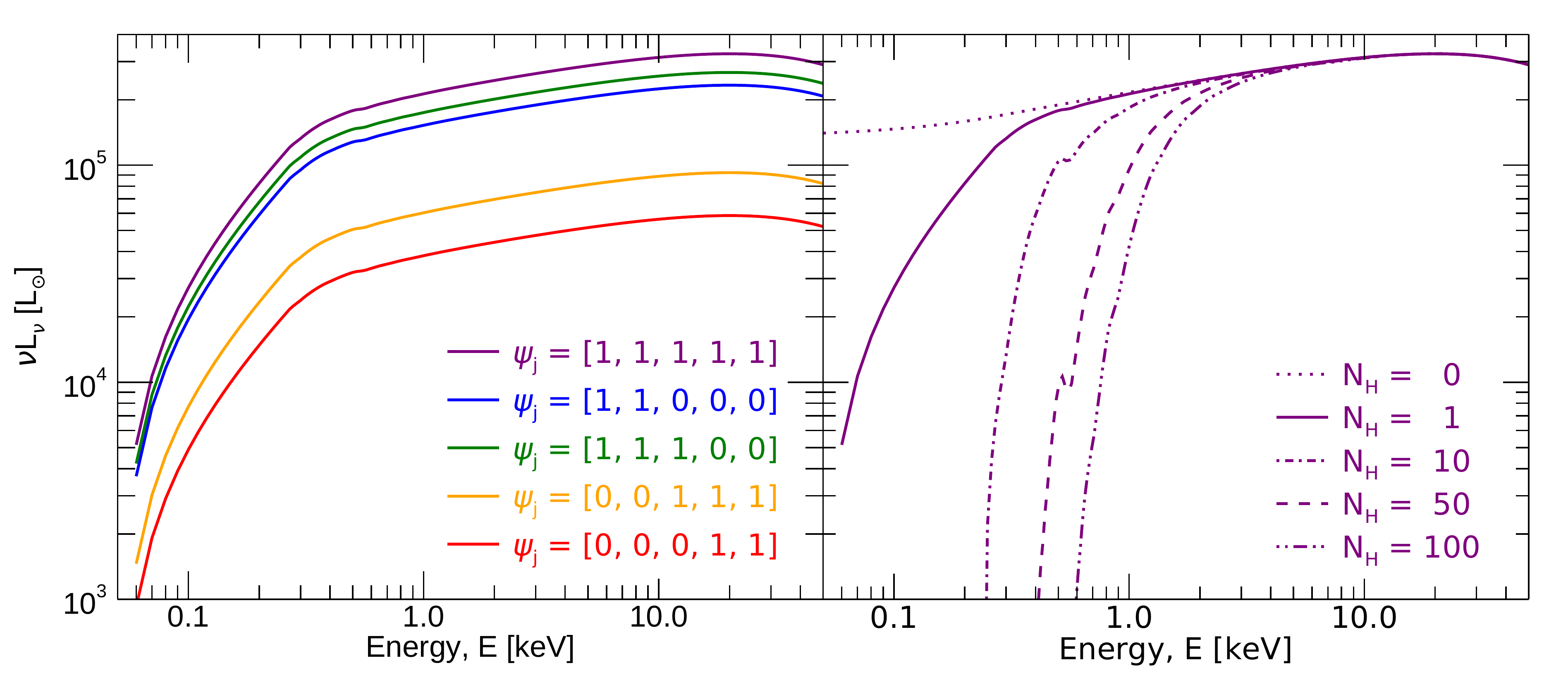}
\caption{
Example X-ray stellar spectra with and without X-ray absorption generated by \texttt{Lightning}. The used absorption model is the \texttt{tbabs} model with \citet{2000ApJ...542..914W} abundances. The variation in color in the left panel corresponds to a change in the SFH parameters $\psi_j$ with fixed $N_{\rm H} = 1$ ($\times 10^{20}\ {\rm cm^{-2}}$). (The age bins for these $\psi_j$ values are the default set of age bins in \texttt{Lightning} as discussed in Section~\ref{sec:SFH}.) The variation in lines styles in the right panel correspond to changes in $N_{\rm H}$ with fixed $\psi_j = [1, 1, 1, 1, 1]\ M_\odot\ {\rm yr}^{-1}$. The blue line in the left panel shows an example of a spectrum dominated by young stars (i.e. HMXBs), while the red line shows a spectrum dominated by old stars (i.e., LMXBs).
}
\label{fig:xrb}
\end{figure*}

In \texttt{Lightning}, the X-ray absorption is modeled using one of two user-selected X-ray absorption models: the \texttt{tbabs} model with the default \citet{2000ApJ...542..914W} abundances or the \texttt{Sherpa} \texttt{atten} model which combines cross-sections from \citet{1983ApJ...270..119M} and \citet{1994AJ....107.2108R}. The tabulated curves used in \texttt{Lightning} were generated with \texttt{Sherpa}\footnote{\url{https://cxc.cfa.harvard.edu/sherpa/}} v4.13, and normalized to a line-of-sight HI column density of $N_{\rm H} = 10^{20}\ {\rm cm^{-2}}$. At energies larger than 10 keV, these models produce negligible absorption; for convenience we set the optical depth to 0 at $>12$ keV.

The chosen X-ray absorption is first applied to the SED model in the observed frame to account for Galactic absorption by the Milky Way, with the Galactic $N_{\rm H}$ being a user input for each galaxy. Further intrinsic absorption is then applied in the rest frame on the stellar binary population and the AGN emission model, if applicable. If there is no X-ray AGN emission, the $N_{\rm H}$ of the stellar population becomes a free parameter in the model. We show some examples of the X-ray stellar model with varying levels of absorption and SFHs in Figure~\ref{fig:xrb}. The inclusion of X-ray absorption primarily impacts lower energy photons as higher energy X-rays are less likely to be absorbed, with the intensity of the absorption increasing with $N_{\rm H}$.

However, when the model includes an X-ray AGN component, absorption of the stellar X-ray emission is not a free parameter, and is instead linked to the $V$-band attenuation via
\begin{equation}
    N_{\rm H} = (22.4 \times 10^{20}\ {\rm cm}^{-2})\frac{2.5\ \tau_{{\rm DIFF},V}}{\ln(10)}.
\end{equation}
This scaling was chosen to be the average of observed Milky Way $N_{\rm H} - A_V$ relationships \citep{1995A&A...293..889P, 2012ApJ...759...95N}. In this case, the $N_{\rm H}$ value of the nuclear region becomes the free parameter in the model, as we expect the X-ray emission from the AGN to be the dominant component of the X-ray spectrum in most cases.\footnote{If the value of $N_{\rm H}$ is fixed by the user (if, e.g., a previous, reliable measurement is available, or fitting only hard X-ray fluxes to ignore the effects of absorption) when using the power-law AGN or X-ray binary models, the shape of the X-ray spectrum will be completely fixed. However, in the current implementation of the code, the complete X-ray spectrum will still be constructed and integrated over the bandpass(es) at each model evaluation.}

We note here that while $N_{\rm H}$ is allowed to increase above $10^{24}\ {\rm cm^{-2}}$ in our implementation, our X-ray emission models are not currently appropriate for Compton-thick sources, which typically require reflection components that are not included in \texttt{Lightning}.

\subsection{Dust Emission Models} \label{sec:DustEmission}

We model IR dust emission in \texttt{Lightning} using the \citet{2007ApJ...657..810D} model. To briefly summarize, the model details how the dust mass, $M_{\rm dust}$, is exposed to a radiation field intensity, $U$. Analytically, this is given by Equation 23 in \citet{2007ApJ...657..810D},
\begin{align}
  \frac{dM_{\rm dust}}{dU} = & (1-\gamma)M_{\rm dust} \delta(U-U_{\rm min})  \notag\\
    & +\gamma M_{\rm dust} \frac{(\alpha-1)}{(U_{\rm min}^{1-\alpha}-U_{\rm max}^{1-\alpha})}U^{-\alpha}, \ \ \alpha \neq 1,
\end{align}
where the first additive component is a delta function at the minimum radiation field intensity $U_{\rm min}$ and the second component is a power-law of slope $\alpha$ derived between $U_{\rm min}$ and $U_{\rm max}$, the maximum radiation field intensity. The parameter $\gamma$ in each additive component dictates the fraction of the dust mass exposed to the power law compared to the delta function. Additionally, the polycyclic aromatic hydrocarbon (PAH) index, $q_{\rm PAH}$, is a hidden parameter in the model and defines the strength of the PAH emission.

Rather than scaling the model with $M_{\rm dust}$, we instead use the total integrated IR (TIR) luminosity $L_{\rm TIR}$ (i.e., the bolometric luminosity of the dust model), which is proportional to $M_{\rm dust}$. The reason we make this substitution is because of the mechanism generating the dust emission. Simply put, some fraction of UV-to-NIR emission in a galaxy is attenuated by dust. This attenuated energy is conserved via radiation from dust particles at longer wavelengths, mainly across the mid-to-far IR. To account for this conservation of energy, it is expected that the bolometric luminosity of the attenuated light should be equal to the TIR luminosity. This energy conservation (often termed ``energy balance'' in the SED fitting community) is optional when fitting with the dust emission model in \texttt{Lightning}. Therefore, when fitting with energy balance, the dust model has five free parameters: $\alpha$, $U_{\rm min}$, $U_{\rm max}$, $\gamma$, and $q_{\rm PAH}$; and when energy balance is not assumed, $L_{\rm TIR}$ becomes an additional free parameter used for normalization.
\begin{figure*}[t!]
\centering
\includegraphics[width=18cm]{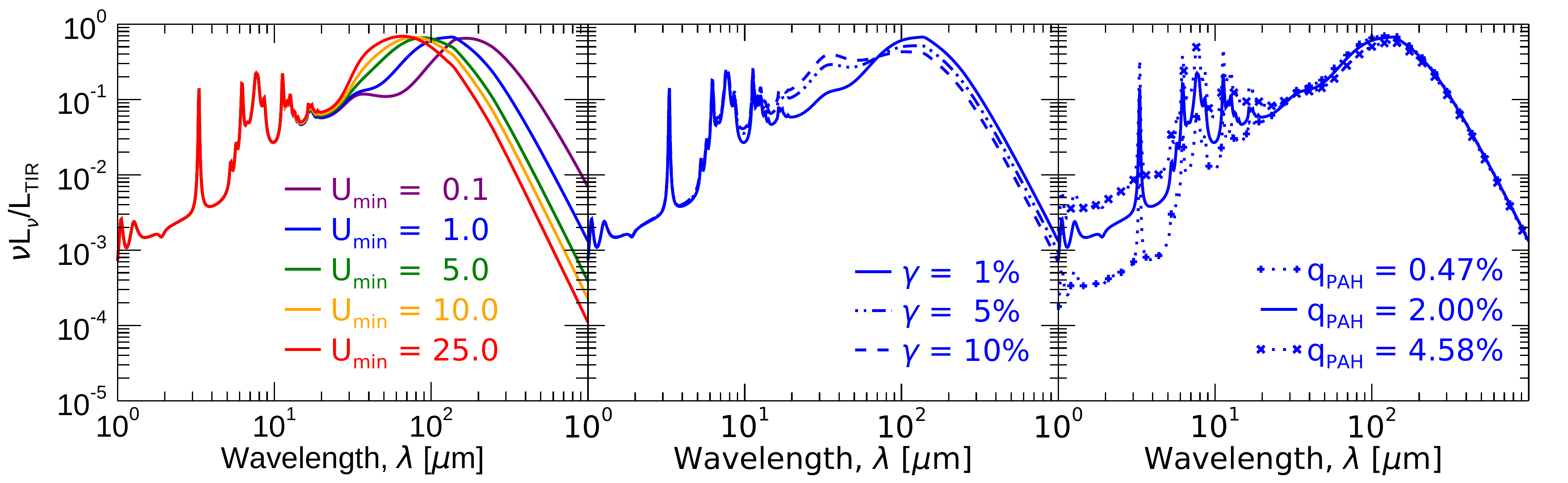}
\caption{
Examples of the \citet{2007ApJ...657..810D} dust emission model generated by \texttt{Lightning}, normalized by $L_{\rm TIR}$. The parameters $U_{\rm max}$ and $\alpha$ are fixed to $3 \times 10^5$ and $2$, respectively. The variation in color in the left panel corresponds to a change in $U_{\rm min}$ with fixed $\gamma = 0.01$ and $q_{\rm PAH} = 0.02$. The different lines styles in the middle panel correspond to variations in $\gamma$ with fixed $U_{\rm min} = 1$ and $q_{\rm PAH} = 0.02$. As for the right panel, the different symbols on the dotted lines correspond to different values of $q_{\rm PAH}$ with fixed $U_{\rm min} = 1$ and $\gamma = 0.01$. Decreasing the value of $U_{\rm min}$ can be seen to shift the peak of the emission to shorter wavelengths, while $q_{\rm PAH}$ independently increases the intensity of the PAH emission features.
}
\label{fig:DustEmis}
\end{figure*}

In Figure~\ref{fig:DustEmis}, we show some examples of the \citet{2007ApJ...657..810D} dust emission model for a variety of input parameters. In these examples and as a default in \texttt{Lightning}, we fix $U_{\rm max}=3 \times 10^5$ and $\alpha = 2$. We make this simplifying assumption as recommended by \cite{2007ApJ...663..866D}, since they found that the dust model is not very sensitive to these two parameters and most observed IR SEDs are well reproduced by $U_{\rm max} = 10^6$ and $\alpha=2$. We note the discrepancy between the fixed value of $U_{\rm min}$ in \texttt{Lightning} and that recommended by \cite{2007ApJ...663..866D}. The reason for the discrepancy comes from how \texttt{Lightning} computes the dust emission model from the publicly available data. To allow for a variable $\alpha$, the $\delta$-functions of $U$ must be used. However, the largest available $\delta$-function is $U=3 \times 10^5$. Therefore, $U_{\rm max}$ is limited to this largest available value, since extrapolating to $U=10^6$ would add unwanted uncertainty to the model.

\section{Statistical Inferencing of SEDs}  \label{sec:Inference}

\subsection{Observational Information}

To keep \texttt{Lightning} computationally fast, we restrict it to modeling only photometric observations, since spectroscopic observations would require additional modeling assumptions. However, any combination of narrow- to broadband flux densities that have been corrected for Galactic extinction can be used to make up the UV-to-submillimeter input SED. Additionally, any number of uniform sensitivity top-hat energy bands can be used for X-ray observations, whose measurements can be in the form of either net counts or fluxes.

Since the models in \texttt{Lightning} are in rest-frame luminosity units, an observed distance indicator is required to convert the SED fluxes to the same luminosity units as the model. The distance indicator can simply be a luminosity distance, or it can be a redshift, where the luminosity distance is calculated from the redshift via the user chosen cosmology. We note that when using a redshift that the assumed age of the Universe, $t_{\rm age}(z)$, will decrease as redshift increases. To account for any effect this will have on the SFH age bins, we have designed \texttt{Lightning} to automatically adjust the user-defined age bins such that upper bin boundaries of $t_{j+1} > t_{\rm age}(z)$ will be updated to $t_{j+1} = t_{\rm age}(z)$, and bins with lower bin boundaries of $t_{j} > t_{\rm age}(z)$ are completely omitted from the SFH.

\subsection{Loss Function} \label{sec:LossFunc}

In order to fit a given model to any data, a loss function, which determines how well the model fits the data, is required. In \texttt{Lightning}, we implement a $\chi^2$ loss function given by
\begin{equation} \label{eq:chi2}
    - \log(\mathcal{L}) \propto \chi^2 = \sum_{f=1}^{n} \frac{(L_{\nu, f}^{\rm obs} - \bar{L}_{\nu, f}^{\rm mod})^2}{\sigma_{{\rm total}, f}^2},
\end{equation}
where $\mathcal{L}$ is the likelihood probability of the model, $L_{\nu, f}^{\rm obs}$ is the observed luminosity in filter $f$ as derived from the input flux, $\bar{L}_{\nu, f}^{\rm mod}$ is the model photometry in filter $f$, and $\sigma_{{\rm total}, f}$ is the total uncertainty associated with filter $f$. The model photometry is derived by integrating the observed-frame model spectrum through the corresponding filter $f$ using
\begin{equation}
    \bar{L}_{\nu, f}^{\rm mod} = \frac{\int T_f(\lambda) L_{\nu}^{\rm mod}(\lambda)\, d\lambda}{\int T_f(\lambda)\, d\lambda},
\end{equation}
where $T_f(\lambda)$ is the filter transmission function in units of energy, and $L_{\nu}^{\rm mod}(\lambda)$ is the model spectrum. The total uncertainty consists of both the observed uncertainty, which is the Gaussian uncertainty of the measurement plus any additional fractional calibration uncertainty, and the model uncertainty added in quadrature or
\begin{equation}
    \sigma_{{\rm total}, f}^2 = \sigma_{{\rm obs}, f}^2 + \sigma_{{\rm mod}, f}^2,
\end{equation}
where $\sigma_{{\rm obs}, f}$ and $\sigma_{{\rm mod}, f}$ are the observed and model uncertainties, respectively. We include a model uncertainty component in \texttt{Lightning} to account for any oversimplification of models and potential systematic uncertainties in the models themselves \citep{1996ApJ...457..625C, 2009ApJ...703.1123P, 2009ApJ...699..486C, 2010ApJ...708...58C, 2010ApJ...712..833C}. The model uncertainty is computed simply as a user-defined fraction of the model photometry that is constant for each filter or
\begin{equation}
    \sigma_{{\rm mod}, f}^2 = (\sigma_{\rm mod}^{\rm frac} \times \bar{L}_{\nu, f}^{\rm mod})^2
\end{equation}
where $\sigma_{\rm mod}^{\rm frac}$ is the user-defined fractional model uncertainty. Therefore, by including model uncertainties, our formulation of the $\chi^2$ loss function accounts for both the observational uncertainty and the inherent uncertainty of the models being used.

When fitting X-ray data in terms of counts, we note that Equation~\ref{eq:chi2} will not be applicable, since the Poissonian nature of the counts is not compatible with uncertainty formulation. Instead, we calculate the $\chi^2$ contribution of the X-ray counts as
\begin{equation}
    \chi_{\rm X}^2 = \sum_{e=1}^{n} \frac{(N_{e}^{\rm obs} - N_{e}^{\rm mod})^2}{\sigma_{N, e}^2},
\end{equation}
where $\chi_{\rm X}^2$ is X-ray counts contribution to the total $\chi^2$, $N_{e}^{\rm obs}$ is the observed net counts (i.e., background subtracted) in energy band $e$, $N_{e}^{\rm mod}$ is the model net counts in energy band $e$, and $\sigma_{N, e}$ is the approximate count uncertainty in energy band $e$. Since the approximate count uncertainty can be computed in a variety of ways depending on the overall count rate, we allow for the user to either input their own pre-computed count uncertainties or use one of the two built-in methods in \texttt{Lightning}. The first method simply sets the count uncertainty equal to the square root of the counts (i.e., $\sigma_{N, e} = \sqrt{N_{e}^{\rm obs}}$), since this is the Gaussian approximation in the high count regime. The other uses the upper uncertainty of the \citet{1986ApJ...303..336G} approximation given by
\begin{equation}
    \sigma_{N, e} = 1 + \sqrt{0.75 + N_{e}^{\rm obs}},
\end{equation}
which is more appropriate for data in the low count regime. User-supplied uncertainties may be used when more flexibility is required, e.g., when the background contributes strongly to the uncertainty.

\subsection{Maximum-Likelihood Inferencing} \label{sec:MaxLike}

\subsubsection{MPFIT Algorithm} \label{sec:MPFIT}

To allow for the determination of a rapid best-fit solution to an SED without sampling the parameter posteriors, we have added a new maximum-likelihood inferencing method to \texttt{Lightning}. The method utilizes the \texttt{MPFIT} code \citep{1978LNM...630..105M, 2009ASPC..411..251M}, which consists of the gradient-descent Levenberg–Marquardt algorithm used to solve nonlinear least-squares problems. We chose the \texttt{MPFIT} implementation since it allows for several necessary constraints in \texttt{Lightning}, such as fixing parameters and setting parameter bounds. 

When searching for the best solution, one drawback to gradient-decent algorithms like \texttt{MPFIT} is that they are prone to getting stuck in local minima in $\chi^2$ space before reaching the global minimum. To mitigate this drawback, we have implemented the algorithm to run a user-defined number of ``solvers'', where each solver is a fit to the SED using different starting locations in parameter space (see Section~\ref{sec:AlgInit}). By running multiple solvers from different starting locations, we can compare the solver solutions. If the majority of the solvers have converged to the same solution, then we can be confident that this is likely the best solution and global minimum. More specifically, we require that (1) at least 50\% of the solvers have $\chi^2 - 4 \leq \chi^2_{\rm best}$, where $\chi^2_{\rm best}$ is the minimum $\chi^2$ value of all solvers, and (2) all of these solvers satisfying criterion (1) have free parameter values that are within a 1\% difference of the best fit.\footnote{The chosen difference in $\chi^2$ of 4 and the 1\% difference are arbitrarily chosen from our test fits to well behaved SEDs. We include the $\chi^2$ result and the parameter values for each solver in the output such that a user can further evaluate these cutoffs at their discretion.} Finally, once convergence to the best solution has been confirmed, the best-fit solver is considered the solution to the SED fit.

\subsection{Bayesian Inferencing} \label{sec:Bayes}

To allow for a high quality sampling of the posterior to an SED model, we implement two Bayesian inferencing methods in \texttt{Lightning}. Both methods utilize MCMC algorithms to sample the posterior distributions of the free parameters, while incorporating prior distributions of these parameters.

\subsubsection{Prior Distributions}

In terms of analytical priors, we only include two basic options in \texttt{Lightning}: truncated uniform and normal (Gaussian) priors. We implement by default some truncated priors, since practically all possible free parameters have at least a lower bound. Therefore, limiting the prior range to comply with the physically allowed values is required. Additionally, both analytical priors are implemented in each parameter's sampled space (e.g., a parameter sampled in log space has a uniform or normal prior in log space). 

Besides the analytical priors, we include the option for a user to input a prior of any shape in tabulated form. The only restriction for the shape of these priors is that they conform to a parameter's physically allowed range. Otherwise, any shape is allowed, which can be useful for creating complex prior distributions for a given parameter. However, we do note that no free parameters can be linked together via the prior, tabulated or analytical. We exclude user-specified parameter linking in \texttt{Lightning}, which is common in other SED fitting codes, to minimize the computational complexity and increase computational speed. Additionally, we automatically link commonly correlated parameters (e.g., $\tau_{{\rm DIFF}, V}$ and $N_{\rm H}$), which would potentially be linked by the user.

With the priors specified, the posterior probability is calculated using
\begin{equation}
    \log(P_{\rm post}) \propto \log(P_{\rm prior}) + \log(\mathcal{L}),
\end{equation}
where $P_{\rm post}$ is the posterior probability and $P_{\rm prior}$ is the prior probability.\footnote{The $\mathcal{L}$ here includes the contributions from both $\chi^2$ (UV-to-IR) and $\chi^2_{\rm X}$ (X-ray).} Since the goal of the MCMC algorithms is to sample $P_{\rm post}$, the loss function from Section~\ref{sec:LossFunc} is updated for these algorithms to include the prior information such that we are minimizing and sampling $-\log(P_{\rm post})$ space rather than $\chi^2$ space.

\subsubsection{Adaptive MCMC Algorithm}

The original MCMC algorithm adopted in \texttt{Lightning} was implemented and discussed in \citet{2021ApJ...923...26D}. The algorithm is an adaptive version of the standard Metropolis–Hastings algorithm \citep{Metropolis1953, Hastings1970} created by \citet{Andrieu2008}. The algorithm simply adjusts the proposal density distribution to achieve an optimal acceptance ratio. Additionally, this adjustment of the proposal density is vanishing, meaning the adaptiveness decreases with each subsequent iteration. Therefore, after many iterations the adaptiveness is insignificant and the algorithm is practically equivalent to the standard Metropolis–Hastings algorithm.

Similar to the \texttt{MPFIT} algorithm, a single chain of the adaptive MCMC algorithm is prone to getting stuck in local minima in $-\log(P_{\rm post})$ space. Once stuck, it can take more than the user-specified number of trials to escape and move towards the global minimum.\footnote{All MCMC algorithms by design will reach and sample the global minimum in $-\log(P_{\rm post})$ space. However, this may take an arbitrarily large number of trials.} To confirm if a chain reached the vicinity of the global minimum, we have designed the adaptive MCMC algorithm to run a user-defined number of independent chains in parallel, where each chain is initialized using different locations in parameters space (see Section~\ref{sec:AlgInit}). By running parallel chains from different starting locations, we can compare the ending segment of each chain to see if they converged to a single best solution. To test for convergence, we have \texttt{Lightning} automatically perform the Gelman-Rubin test \citep{Gelman1992, Brooks1998} and its multivariate version from \citet{Brooks1998} on the ending segment of the chains, whose length is user-defined. If the ending segments from the parallel chains results in acceptable Gelman-Rubin statistics (i.e., $\sqrt{\hat{R}} \leq 1.2$), then a user can confidently concluded that convergence has been reached and the posterior well sampled by the algorithm.

Since the ending portion of the full parallel chains, which samples the posterior, is the main interest of a user, \texttt{Lightning} automatically post-processes the full chains to create a final post-processed chain portion. For the adaptive MCMC algorithm, this post-processed chain portion is determined as follows. First, each parallel chain has a user-defined number of initial trials discarded as the burn-in phase. Next, these truncated chains are thinned by a user-specified thinning factor (i.e., only every $n$ elements of each chain is kept). Finally, the thinned and truncated chain containing the highest posterior probability is selected, and the ending segment of this highest probability chain, whose length is user-defined, is kept as the sampled posterior. 

The reason for selecting the highest posterior probability chain is based on the assumption that convergence may not have been reached. If convergence was reached, then all parallel chains will have very similar distributions, and it does not matter which one is selected for use. However, if convergence was not reached, selecting the chain with the highest probability, guarantees that the chain with the best solution is selected.

\subsubsection{Affine-Invariant MCMC Algorithm} \label{sec:Affine}

To more quickly sample the potentially skewed posteriors of the free parameters, we have recently added an affine-invariant MCMC algorithm to \texttt{Lightning} \citep{2023ApJ...951...15M}, which is our default algorithm. This algorithm uses an ensemble of samplers to adjust the proposal density distribution and sample the posterior distribution. The ensemble consists of multiple chains (or walkers) that are run in parallel and allowed to interact with one another so that they can adapt their proposal densities. For our implementation in \texttt{Lightning}, we use the affine-invariant ``stretch move'' method as presented in \citet{2010CAMCS...5...65G}, which was shown to more quickly sample non-Gaussian and skewed posteriors compared to the Metropolis–Hastings MCMC algorithms.

Unlike the \texttt{MPFIT} and adaptive MCMC algorithms, the ensemble nature of the affine-invariant MCMC algorithm typically prevents it from getting stuck in local minima in $-\log(P_{\rm post})$ space, since each walker is initialized using different locations in parameters space (see Section~\ref{sec:AlgInit}). However, it is still important to confirm that the ensemble has converged to a stationary solution and to
quantify any potential sampling error effects, since each walker in the ensemble is not independent. To test for this convergence, we have \texttt{Lightning} automatically perform an autocorrelation analysis on the ensemble (see \citealt{2010CAMCS...5...65G} and \citealt{2013PASP..125..306F} for detailed discussions on autocorrelation analyses). We briefly summarize the idea and methods of the analysis as applied in \texttt{Lightning} below.

Autocorrelation, in the context of MCMC algorithms, is how correlated a sample is with previous samples from the same walker or chain. Specifically in \texttt{Lightning}, we look at the integrated autocorrelation time, which is a measure of the average number of iterations between independent samples. If the autocorrelation time is large, then the samples in the ensemble are likely highly correlated and contain few independent samples. To confirm the ensemble has enough independent samples and to quantify the Monte Carlo error, we have designed \texttt{Lightning} to calculate the autocorrelation time and check for convergence following the methods of \citet{2013PASP..125..306F}. Their methods recommend that the MCMC algorithm runs for a number of iterations at least $\sim$50 times the integrated autocorrelation time in order for one to trust that the autocorrelation time estimate is accurate and convergence has been reached. A factor fewer than $\sim$50 can cause the autocorrelation time to be underestimated, which could result in a highly correlated sampling with few independent samples. Therefore, we have \texttt{Lightning} check if the user-defined number of iterations is large enough to have an accurate autocorrelation time estimate (i.e., autocorrelation time $\ge$50) and flag fits that do not.

Similar to the adaptive MCMC, the ensemble is automatically post-processed by \texttt{Lightning} to produce a final post-processed chain portion. For the affine-invariant MCMC algorithm, the post-processed chain portion is determined as follows. First, each walker in the ensemble has its burn-in phase discarded, where the burn-in phase is a number of iterations from the beginning of the chain equal to either twice the longest autocorrelation time of any parameter in the ensemble (double the autocorrelation time typically encapsulates the entire burn-in phase) or a user-defined value. Next, the truncated ensemble is then thinned by a thinning factor, which is either half the longest autocorrelation time of any parameter in the ensemble (iterations at half the autocorrelation time typically give fully independent samples) or user-defined. Then, if a walker in the thinned and truncated ensemble is classified as stranded (we explain how we classify walkers as stranded below), they are removed from the ensemble. Finally, the nonstranded ensemble is flattened element-wise into a single chain and the ending segment of this flattened chain, whose length is user-defined, is kept as the sampled posterior. 

We designed \texttt{Lightning} to classify walkers as stranded if they have an acceptance fraction less than a user-specified number of standard deviations below the median ensemble acceptance fraction. Due to the boundaries of the free parameters, the affine-invariant MCMC can have trouble accepting moves of walkers separated from the ensemble (typically at higher $-\log(P_{\rm post})$ values) when the ensemble is near a boundary. This results in the walkers becoming stranded and having very low acceptance rates, since they are failing to have any proposal jumps accepted. With enough iterations, these walkers will eventually have a jump that rejoins them with the ensemble. However, only a finite number of iterations are allowed for this to occur. Therefore, once the specified iteration limit has been reached, any stranded walkers that may remain need to be removed, since they would add faulty samples to the final sampled posterior. We have found that the most effective automatic method for correctly selecting stranded walkers is to compare each walker's acceptance fraction with that of the median of the ensemble. Those that are classified as stranded with abnormally low acceptance fractions compared to the rest of the ensemble are consistently considered stranded when using more robust and manual visualization methods.

\subsection{Algorithm Initialization} \label{sec:AlgInit}

All three of the current algorithms in \texttt{Lightning} require initial starting values for each free parameter. To select these starting values, \texttt{Lightning} randomly samples the user-specified prior distribution of each parameter independently. (For the \texttt{MPFIT} algorithm, there are technically no priors, since it is not a Bayesian inferencing method. However, we assume uniform ``priors'' for all parameters for the purpose of initialization.) Additionally, since a given prior may have a much larger range than what would constitute an appropriate starting range, we allow the user to limit the initialization to a specified range within the prior range. Therefore, each initialization of the algorithms (i.e., each unique solver, chain, or walker) is initialized randomly from a potentially limited range of the corresponding prior.

\subsection{Derived Quantities} \label{sec:DerivedQuant}

\begin{figure*}[t!]
\centering
\includegraphics[width=18cm]{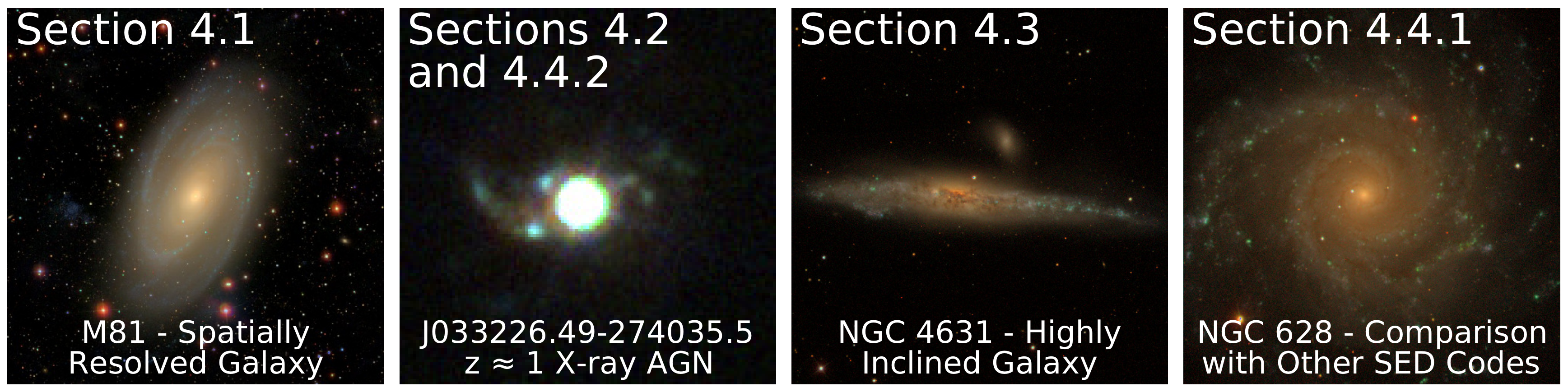}
\caption{
Composite SDSS $g$, $r$, $i$ (M81, NGC~4631, and NGC~628) and HST F435W, F606W, and F850LP (J033226.49$-$274035.5) postage stamp images for the galaxies used in the example applications of \texttt{Lightning} in Section~\ref{sec:Examples}.
}
\label{fig:ExampleTOC}
\end{figure*}

After fitting an SED with the chosen algorithm, we designed \texttt{Lightning} to automatically do additional post-processing to derived typical quantities of interest. In terms of physical properties that are not free parameters in the model, the total and individual model component spectra and photometry with and without attenuation are derived. When using an MCMC algorithm, \texttt{Lightning} allows for a user to select and keep the model spectra for a specified fraction of best-fit elements in the final post-processed chain portion. This is useful for deriving and quoting model uncertainties on new simulated photometric data points as well as showing model uncertainties in the spectra when plotting. Additional physical properties are also derived, such as the stellar mass and $L_{\rm TIR}$, if their corresponding model component is included in the total model.

One other important quantity of interest that \texttt{Lightning} automatically derives is a $p$-value from a goodness-of-fit test for each SED fit. Goodness-of-fit tests are often overlooked in SED fitting, but they are important for determining if the chosen model can acceptably model the data and whether the uncertainties are trustworthy. For the \texttt{MPFIT} algorithm, we use a $\chi^2$ goodness-of-fit test to derive the $p$-value using the $\chi^2$ and degrees of freedom as given by the \texttt{MPFIT} algorithm. We caution against using this $p$-value to reject the null hypothesis that the chosen model can acceptably model the given SED. Since the effective number of free parameters is lower than the actual number (due to degeneracies and covariances between parameters), the number of degrees of freedom is likely higher than what is given by \texttt{MPFIT}. Therefore, this $p$-value can be underestimated and can lead one to falsely conclude that the model is not acceptable for the given SED.

As for the MCMC algorithms, since they are Bayesian methods, \texttt{Lightning} performs a posterior predictive check \citep[PPC;][]{Rubin1984, Gelman1996} to derive a $p$-value for the chosen model. A PPC is a goodness-of-fit test that uses the model itself to estimate the distribution from which the $p$-value is derived. The model can be considered an accurate description of the data if it can generate simulated (replicated) data that is statistically identical to the actual data. By considering the replicated data as data that could have been measured, the PPC tests whether the model encapsulates all of the variability in the actual data. 

In terms of the practical application in \texttt{Lightning}, a PPC takes the following steps. First, replicated sets of model photometry are randomly selected from the posterior distribution (i.e., samples are bootstrapped from the post-processed chain portion with the chance of selection being weighted by their posterior probability). Then, the replicated sets of photometry are randomly perturbed by a Gaussian distribution with a variance corresponding to the respective total uncertainty. Next, likelihood probabilities are calculated (see Section~\ref{sec:LossFunc}) by comparing the actual observations and the perturbed replicated photometry with the unperturbed replicated photometry. Finally, the $p$-value is computed as the fraction of corresponding likelihood probabilities for the perturbed replicated photometry that are smaller than the likelihood probabilities for the actual observations.\footnote{See Section~5.1 of \citealt{2016MNRAS.462.1415C} for a more detailed description of how PPCs are used in SED fitting.}

\section{Example Applications}  \label{sec:Examples}

In this section, we present different example applications of \texttt{Lightning} to give interested users ideas of its capabilities. In Figure~\ref{fig:ExampleTOC}, we show composite postage stamp images of the galaxies used in each example, along with a brief description of the example topic. We note that, within the ``examples'' sub-directory of the \texttt{Lightning} GitHub repository and online documentation\footnote{\url{https://lightning-sed.readthedocs.io/en/latest/}}, we provide the scripts required to run the examples and generate their presented figures. For interested users, we recommend following along in the online documentation in addition to the below text to get further supplementary details.

For all of these examples, we adopt a cosmology with $H_0=70~\rm{km~s}^{-1}~\rm{Mpc}^{-1}$, $\Omega_M=0.30$, and $\Omega_{\Lambda}=0.70$.

\subsection{Property Maps of M81}

To show the power and versatility of \texttt{Lightning} when applied to a nearby galaxy, we fit the spatially resolved UV-to-FIR photometry (SED map) of the nearby spiral galaxy, M81 (NGC~3031). To generate the SED map, we gathered 23 publicly-available photometric images ranging in wavelength from GALEX FUV to Herschel 350~$\mu$m. We then preprocessed these images for fitting by (1) subtracting foreground stars, (2) convolving each image to a common 25$^{\prime\prime}$ PSF, (3) re-binning them to a common 10$^{\prime\prime}$ pixel scale,  (4) estimating the background to update the photometric uncertainties, (5) correcting each bandpass for Galactic extinction, and (6) combining the images into a data cube which contains the pixel-by-pixel SEDs. Since any data pre-processing steps are done external to running \texttt{Lightning}, we exclude the intricate details on our image pre-processing methods as the focus of this example is on the application of \texttt{Lightning} rather than the creation of SED maps. Interested readers can refer to Section~2 of \citet{2017ApJ...851...10E} for a detailed description of the pre-processing steps involved.

To model the SEDs, we used a stellar population with solar metallicity (i.e, $Z=0.02$) and SFH age bins of 0--10~Myr, 10--100~Myr, 0.1--1~Gyr, 1--5~Gyr, and 5--13.6~Gyr. The stellar emission was attenuated using the modified \citet{2000ApJ...533..682C} curve with the 2175~\AA\ bump feature and excluding any birth cloud attenuation. Finally, the dust attenuation was set to be in energy balance with the \cite{2007ApJ...657..810D} dust emission model, where we fixed $U_{\rm max}=3 \times 10^5$ and $\alpha=2$, as recommended by \cite{2007ApJ...663..866D} and discussed in Section~\ref{sec:DustEmission}. To fit the model to each SED, we utilized the affine-invariant MCMC algorithm, which we ran for $10^4$ trials with 75 walkers, assuming 5\% model uncertainty. For all free parameters, we implemented uniform priors over either the entire available range or a broad range of values if the available range is infinite. The values defining the priors associated with each free parameter in the model, along with the limited initialization ranges, are listed in Table~\ref{tab:M81Param}. 
\input{table_m81.tex}

\begin{figure*}[t!]
\centering
\includegraphics[width=18cm]{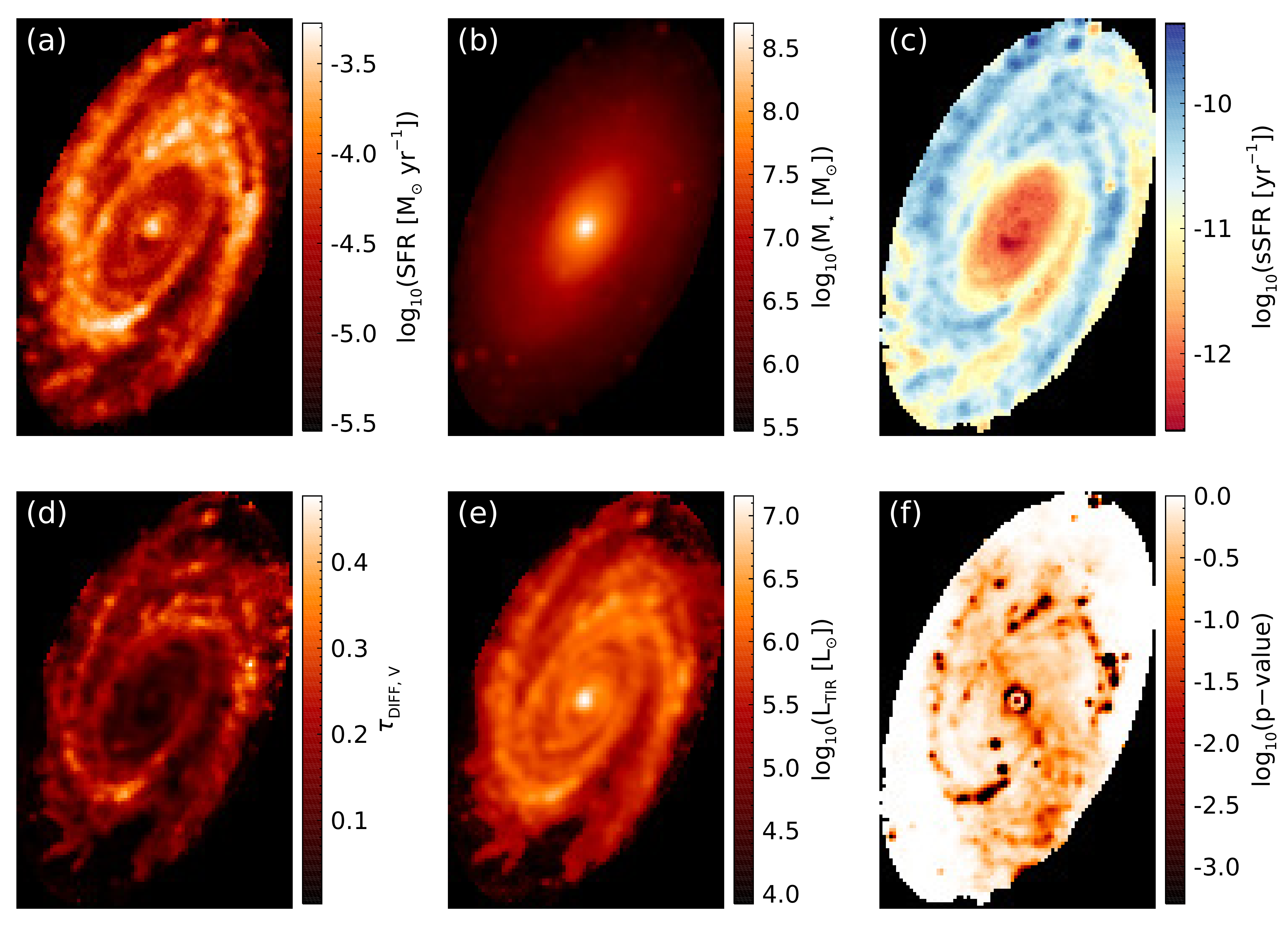}
\caption{
Maps of the derived spatially resolved properties of M81. The values of each pixel are the median value from the corresponding posterior distribution (excluding the $p$-value, which is a single value, in panel (\textit{f})). The properties from the upper left to bottom right are (\textit{a}) the SFR of the last 100 Myr, (\textit{b}) the stellar mass, (\textit{c}) the sSFR of the last 100 Myr, (\textit{d}) the diffuse $V$-band optical depth, (\textit{e}) the $L_{\rm TIR}$, and (\textit{f}) the $p$-value estimated from the PPC. From these maps, it can be seen that the attenuation and recent star formation is concentrated in the spiral arms, while the stellar mass is concentrated in the bulge. This age stratification is clear in the sSFR map, which uses a different color scheme to highlight younger populations in blue and older populations in red.
}
\label{fig:M81Maps}
\end{figure*}

With the described model and algorithm, we used \texttt{Lightning} to fit a subset of the SEDs within the SED map, assuming a luminosity distance to M81 of 3.5~Mpc as given in \citet{2017ApJ...837...90D}. The subset included all SEDs that were inside the 25~mag arcsec$^{-2}$ $B$-band isophotal ellipse as given by HyperLeda\footnote{\url{http://leda.univ-lyon1.fr}} \citep{2014A&A...570A..13M}, which limited our example to the general extent of the optical emission of the galaxy. The fitting process took 65.1 hours total to fit all 6972 SEDs using one 32-core, 2.1 GHz CPU on the Arkansas High Performance Computing Center (i.e., 0.30 core-hours per SED), with \texttt{Lightning} automatically running each SED fit in parallel to maximize CPU usage. To give a general sense of fitting speed, other Bayesian sampling SED fitting codes typically take 1-100 core-hours per SED depending on the complexity of the chosen model, which means \texttt{Lightning} fits $\gtrsim$1 order of magnitude faster than other codes.

Once \texttt{Lightning} is finished fitting, it automatically post-processes the fit to each SED as described in Sections~\ref{sec:Affine} and~\ref{sec:DerivedQuant} and combines the results into a single file. When generating the final post-processed chain portion, we manually set the length of the burn-in phase and thinning factor for consistency rather than having \texttt{Lightning} automatically determine them for each SED from the autocorrelation times. We chose a burn-in length of 8000 trials and a thinning factor of 250, which is significantly larger than the overall maximum autocorrelation time of 178. Therefore, each element in the final chain portions should be uncorrelated. Finally, we only keep the final 250 elements of each chain so that we can derive reasonable median and 16th and 84th percentile ranges, while minimizing the total memory of the single post-processed file.
\begin{figure*}[t!]
\centering
\includegraphics[width=18cm]{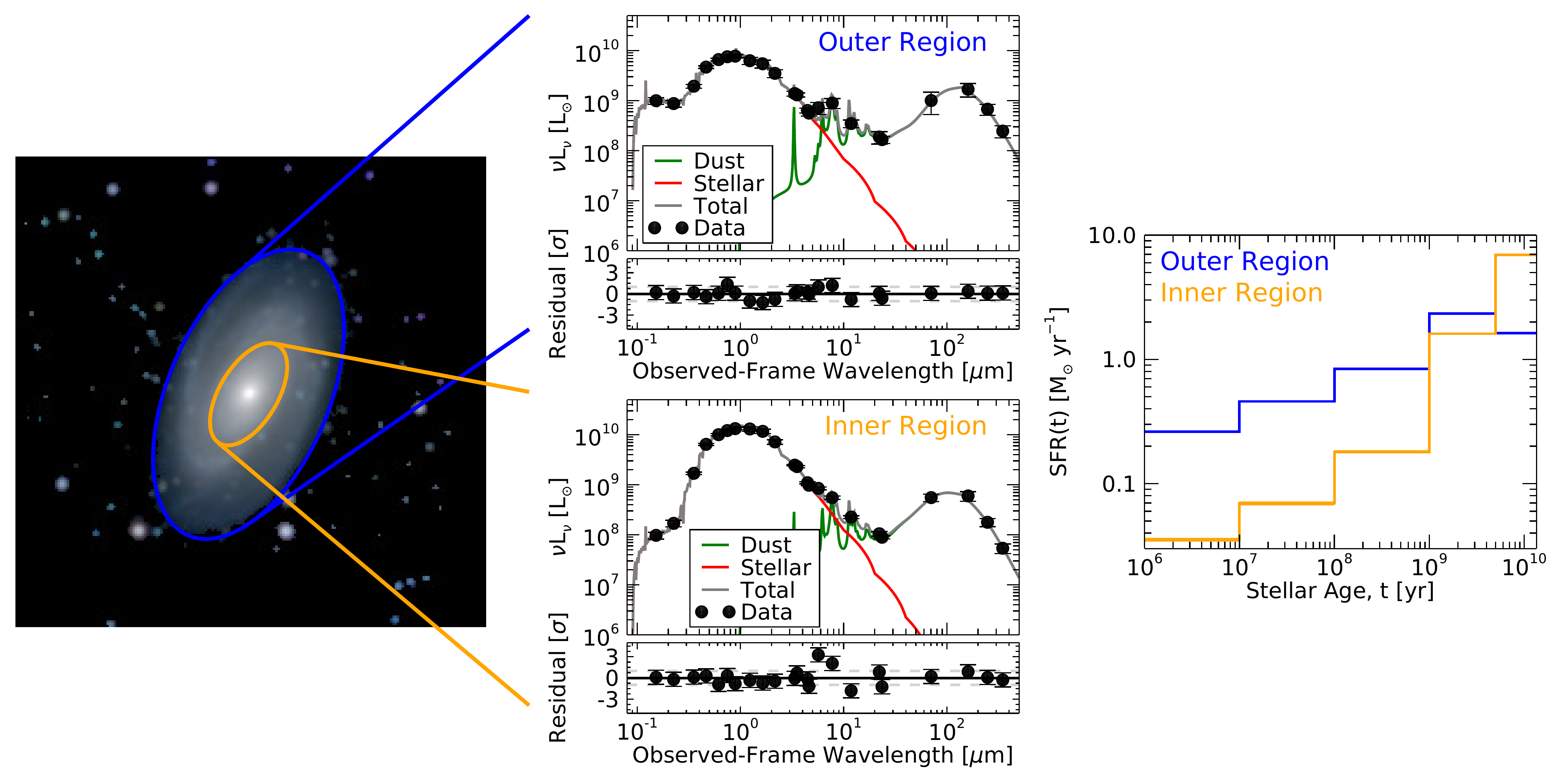}
\caption{
(\textit{Left}) Composite SDSS $g$, $r$, $i$ color image of M81 after being convolved to a common PSF of 25$^{\prime\prime}$ and re-binned to a pixel scale of 10$^{\prime\prime}$. Overlain are the 25~mag arcsec$^{-2}$ $B$-band isophotal ellipse as given by HyperLeda in blue and one half of the 20~mag arcsec$^{-2}$ $K_s$-band isophotal ellipse as given by \citet{2003AJ....125..525J} in orange. (\textit{Middle}) Total best-fit model spectra and component SEDs of the outer (upper plot) and inner (lower plot) regions of the galaxy derived by summing the individual pixels, where the outer region is all pixels between the $B$-band ellipse and the half of the $K_s$-band ellipse (i.e., between blue and orange ellipses) and the inner region is all pixels within the half of the $K_s$-band ellipse (i.e., inside orange ellipse). (\textit{Right}) Resulting SFH with uncertainty ranges for the outer and inner regions as the blue and orange lines, respectively. From the SEDs and SFHs, it can be seen that the outer region has comparatively higher UV emission and lower NIR emission, which is distinguished by the fit as an overall younger population compared to the inner region.
}
\label{fig:M81SEDs}
\end{figure*}

With the post-processed results, we first checked that all SEDs converge to stationary solutions as determined by their autocorrelation times derived from the full chain (see Section~\ref{sec:Affine}). After confirming convergence, we then mapped the derived quantities for each SED back to its associated pixel to generate maps of the spatially resolved properties of the galaxy. In Figure~\ref{fig:M81Maps}, we show a variety of these derived spatially resolved properties, with the values of each pixel being the median value from the corresponding posterior distribution. The only exception to this is the image of the $p$-values in panel (\textit{f}), which only has a single value rather than a distribution. From this image, it can be seen that the vast majority of pixels are well fit by the model, with poor fits mainly being associated with locations where foreground star subtraction occurred. One stand out result from these property maps is that the younger, star-forming population, which is more highly obscured, is concentrated in the spiral arms as seen in the SFR (SFR of the last 100 Myr) and $V$-band optical depth maps. Additionally, the older, more massive population can clearly been seen to reside in the bulge region, where the stellar mass is high and SFR is low, resulting in a low specific star formation rate (sSFR; ${\rm sSFR} = {\rm SFR}/M_\star$).

To further demonstrate how the spatially resolved results could be used to estimate properties for regions of the galaxy, we separate the galaxy into two parts, the outer and inner regions, with the outer region being defined as the pixels between the 25~mag arcsec$^{-2}$ $B$-band isophotal ellipse as given by HyperLeda and one half of the 20~mag arcsec$^{-2}$ $K_s$-band isophotal ellipse as given by \citet{2003AJ....125..525J}, and the inner region comprised of pixels within one half of the 20~mag arcsec$^{-2}$ $K_s$-band isophotal ellipse. In the left of Figure~\ref{fig:M81SEDs}, these regions are shown overlaid on the convolved and re-binned SDSS color image as the blue and orange ellipses, respectively. By summing the results of each pixel within each region, a total SED and SFH can be made as shown in the middle and right of Figure~\ref{fig:M81SEDs}, respectively. These results show, as inferred from the maps in Figure~\ref{fig:M81Maps}, that the outer region has comparatively higher UV emission and lower NIR emission, which is distinguished by the fit as an overall younger population compared to the inner region.
\begin{figure*}[t!]
\centering
\includegraphics[width=18cm]{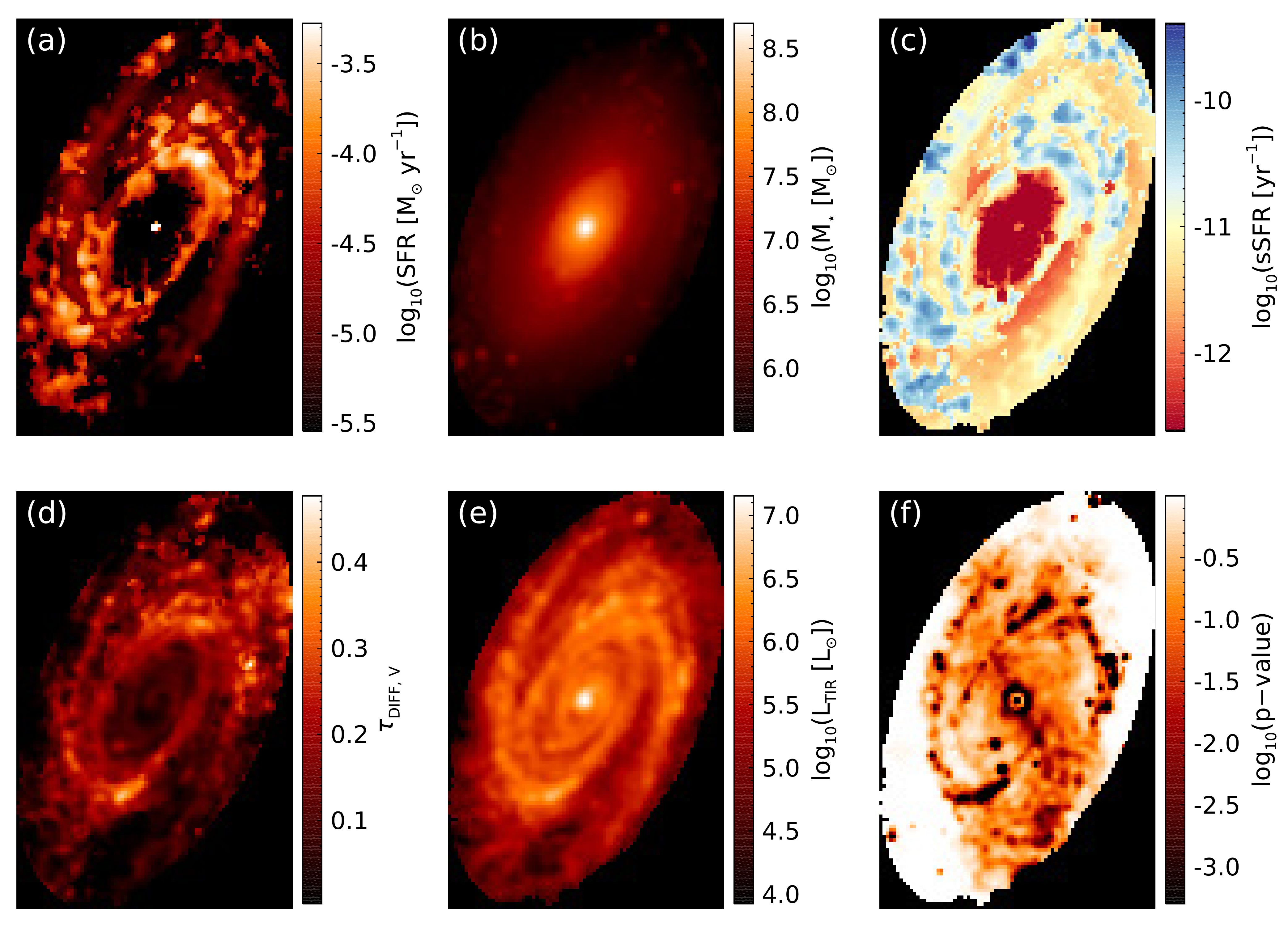}
\caption{
Panels (\textit{a -- e}) are the same as Figure~\ref{fig:M81Maps} except the values correspond to the best-fit values derived from the \texttt{MPFIT} algorithm. The $p$-values in panel (\textit{f}) are estimated from the a $\chi^2$ goodness-of-fit test. The colorbar minimum and maximum ranges have been truncated to match those in Figure~\ref{fig:M81Maps}. Comparing these maps to those in Figure~\ref{fig:M81Maps}, it can be seen that some of these maps, particularly the SFR and sSFR, are not nearly as smooth, due to the use of best-fit values rather than the posterior median.
}
\label{fig:M81MapsMPFIT}
\end{figure*}

Finally, to give a sense of how the maximum likelihood and Bayesian algorithms in \texttt{Lightning} compare, we refit the $B$-band isophote subset of the SED map after swapping the MCMC algorithm with the \texttt{MPFIT} algorithm. Using 20 solvers to test for convergence, fitting took 19.1 hours to fit all 6972 SEDs on one of the same 32-core, 2.1 GHz CPUs (i.e., 0.09 core-hours per SED), which is 3.4 times faster than the MCMC algorithm. In Figure~\ref{fig:M81MapsMPFIT}, we show the same property maps as in Figure~\ref{fig:M81Maps}, except for the best fit as determined by \texttt{MPFIT}. As expected, there are variations in the results between algorithms, especially in the smoothness of the SFR and, subsequently, sSFR maps. This loss of smoothness is primarily due to the usage of the best fit rather than medians of the posterior, since the median of the posterior is rarely the same as the best-fit solution. For example, the central region can be seen to have best-fit solutions that are suggesting zero recent star formation (the colorbar is truncated due to the log-scale and to match the colorbar ranges in Figure~\ref{fig:M81Maps}). These zero values from \texttt{MPFIT} in low SFR regions are not unexpected, since SFR posterior distributions from the MCMC algorithm generally provide upper limits, with the maximum probability coinciding with zero. To show how this variation in best fit and median affects the total SFR of the galaxy, each pixel in the SFR map can be summed to give a total SFR. This results in ${\rm SFR} = 0.22\ M_\odot\ {\rm yr}^{-1}$ and ${\rm SFR} = 0.44\ M_\odot\ {\rm yr}^{-1}$ for the \texttt{MPFIT} and MCMC algorithms, respectively, a difference by a factor of two. In contrast, the stellar mass, in general, is better constrained by SED fits, and the best fit is typically closer in value to the median of the posterior. Summing the pixels for the masses gives $M_\star = 4.05 \times 10^{10}\ M_\odot$ and $M_\star = 4.26 \times 10^{10}\ M_\odot$ for the \texttt{MPFIT} and MCMC algorithms, respectively, a difference of only 5\%. Thus, the best-fit mass map from MPFIT appears nearly identical to the median mass map from the MCMC.

\subsection{Deep Field AGN} \label{sec:DeepAGN}

Here, we demonstrate how the AGN module in \texttt{Lightning} can be used with and without X-ray data by fitting the SED of J033226.49$-$274035.5, an X-ray detected AGN in the Chandra Deep Field South at $z=1.03$. For both fits, the UV-to-NIR photometry were retrieved from the \citet{2013ApJS..207...24G} CANDELS catalog, which covers the $U$-band to Spitzer IRAC 8.0~$\mu$m. From this data, we excluded the VLT/VIMOS~$U$ and HST/ACS~F435W bands from our SED, due to potential contamination by broad-line emission, since J033226.49$-$274035.5 is classified from optical spectra as a type 1 AGN \citep{2004ApJS..155..271S}. We additionally included the FIR data (Spitzer MIPS 24~$\mu$m to Herschel SPIRE 250~$\mu$m) from \citet{2019ApJS..243...22B}. We corrected the CANDELS photometry for Galactic extinction using the \citet{1999PASP..111...63F} curve, with $A_V = 0.025$, as retrieved from the IRSA DUST tool\footnote{\url{https://irsa.ipac.caltech.edu/applications/DUST/}}. To retrieve the X-ray products needed for our fits, we queried the Chandra Source Catalog (CSC) by performing a cone search in $1^{\prime\prime}$ around the source position, finding a unique match, 2CXO J033226.4$-$274035, within $0.41^{\prime\prime}$. We utilized the level 3 CSC spectrum and response files for the single deepest ($\approx 163$ ks) observation of the source (ObsID 5019).
To produce the X-ray photometry used for our fits, we subtracted the scaled background from the spectrum and grouped the X-ray counts into 15 log-spaced bins spanning 0.5--6.0 keV using \texttt{Sherpa} v4.13. We chose this energy range and binning such that the SNR in each bin is $>$2.

We modeled the resulting SED for J033226.49$-$274035.5 both with and without an X-ray model. In both cases, we used a stellar population with solar metallicity (i.e., $Z=0.02$) and stellar age bins spanning 0--10~Myr, 10--100~Myr, 0.1--1~Gyr, 1--5~Gyr, and 5--5.6~Gyr (the age of the universe at $z=1.03$). Both models also included the SKIRTOR UV-to-IR AGN model. Simultaneously constraining the viewing angle and optical depth of the torus is difficult, so we simplified the model by setting $\tau_{9.7} = 7$, the middle of the SKIRTOR model's allowed range. Additionally, since J033226.49$-$274035.5 is classified as a type 1 AGN, we implemented a prior on the cosine of the viewing angle to limit our models to a type 1 AGN. For the fit without an X-ray component, we allow the log of the integrated luminosity of the AGN model to vary between 11.0 and 13.0, with a uniform prior. To attenuate the UV-to-NIR emission (both the stellar and AGN emission that escapes the torus; see Section~\ref{sec:skirtor}), we used the base \citet{2000ApJ...533..682C} curve. The dust attenuation was set to be in energy balance with the \cite{2007ApJ...657..810D} dust emission model, with $U_{\rm max}=3 \times 10^5$, $\alpha=2$, and $q_{\rm PAH} = 0.0047$. We fix $q_{\rm PAH}$ to the minimum allowed value for this example, since high-redshift galaxies like J033226.49$-$274035.5 are not expected to have strong PAH emission.

For the X-ray model in this example, \texttt{Lightning} automatically includes a stellar component when using an X-ray model, and we additionally use the \texttt{qsosed} X-ray model for the AGN X-ray emission component. X-ray absorption was modeled using the \texttt{tbabs} model, with \citet{2000ApJ...542..914W} abundances and a Galactic HI column density fixed at $N_{\rm H} = 9.19\times10^{19}\ {\rm cm^{-2}}$, as retrieved using the \verb|prop_colden| tool in \texttt{CIAO}. We fit both models using the affine-invariant MCMC sampler, with an ensemble of 75 walkers running for $4\times10^4$ steps, assuming 10\% model uncertainty. We adjusted the proposal distribution width parameter $a$ to 1.8 to achieve acceptance fractions $>20\%$. The free parameters and associated priors for these fits are summarized in Table~\ref{tab:J033226Param}. We set \texttt{Lightning} to automatically generate the final chain portion of the posterior distributions from the MCMC chains and keep the last 1000 posterior samples, with the autocorrelation times indicating convergence of the runs.
\input{table_J033226}

\begin{figure*}
\centering
\includegraphics[width=15cm]{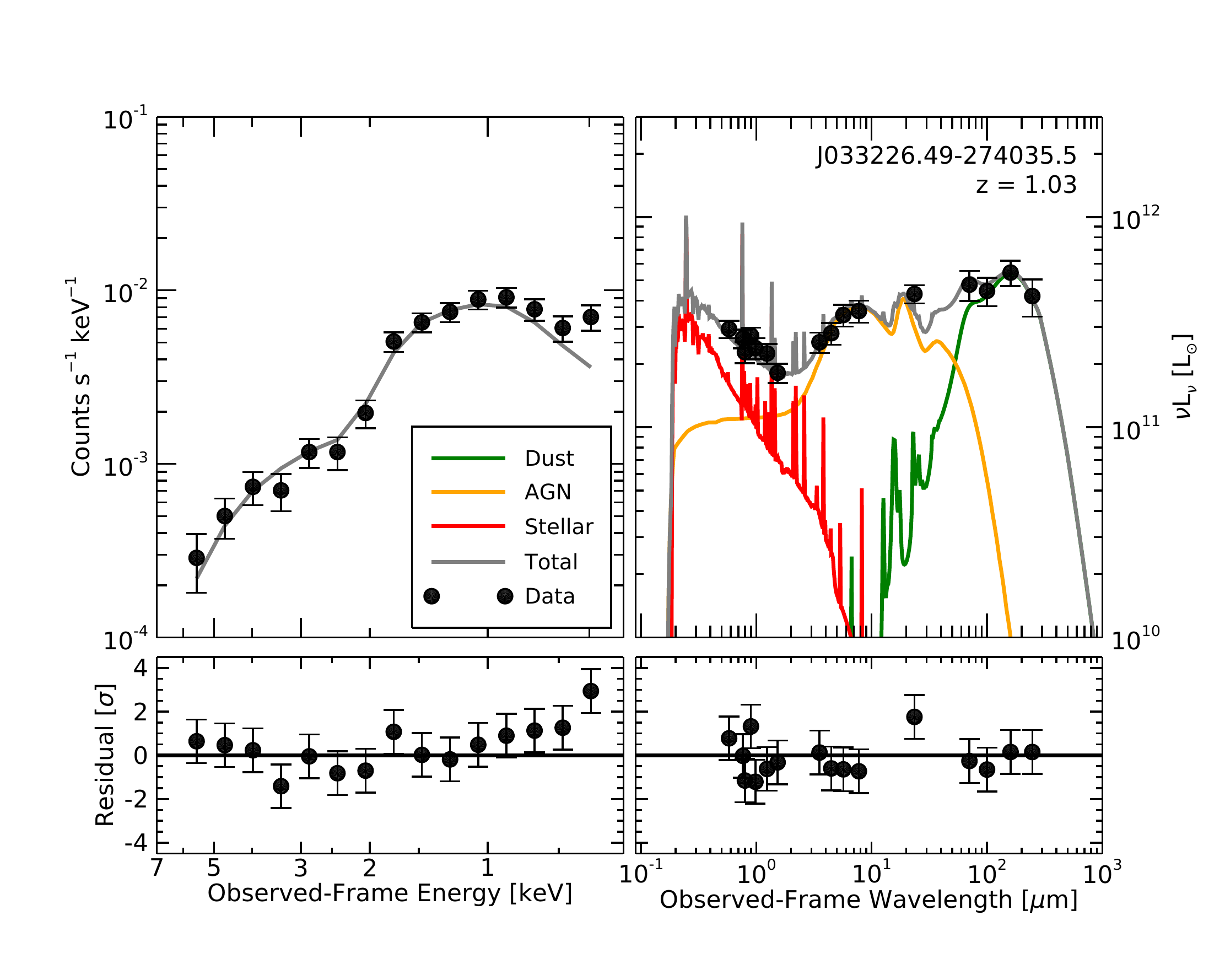}
\caption{
The X-ray-to-IR SED fit for J033226.49$-$274035.5. In the left panels, we show the instrumental X-ray spectrum (in terms of count rate density) with its best-fit model and residuals. In the right panels, we show the observed UV-to-IR SED (in terms of luminosity), its best-fitting model, and residuals. The best-fit model minimizes the total X-ray and UV-to-IR $-\log(P_{\rm post})$. The \texttt{Lightning} X-ray model implementation can provide rudimentary X-ray spectral fits, and directly connect them to the UV-to-IR SED fit.
}
\label{fig:J033226_SEDs}
\end{figure*}

\begin{figure*}
\centering
\includegraphics[width=18cm]{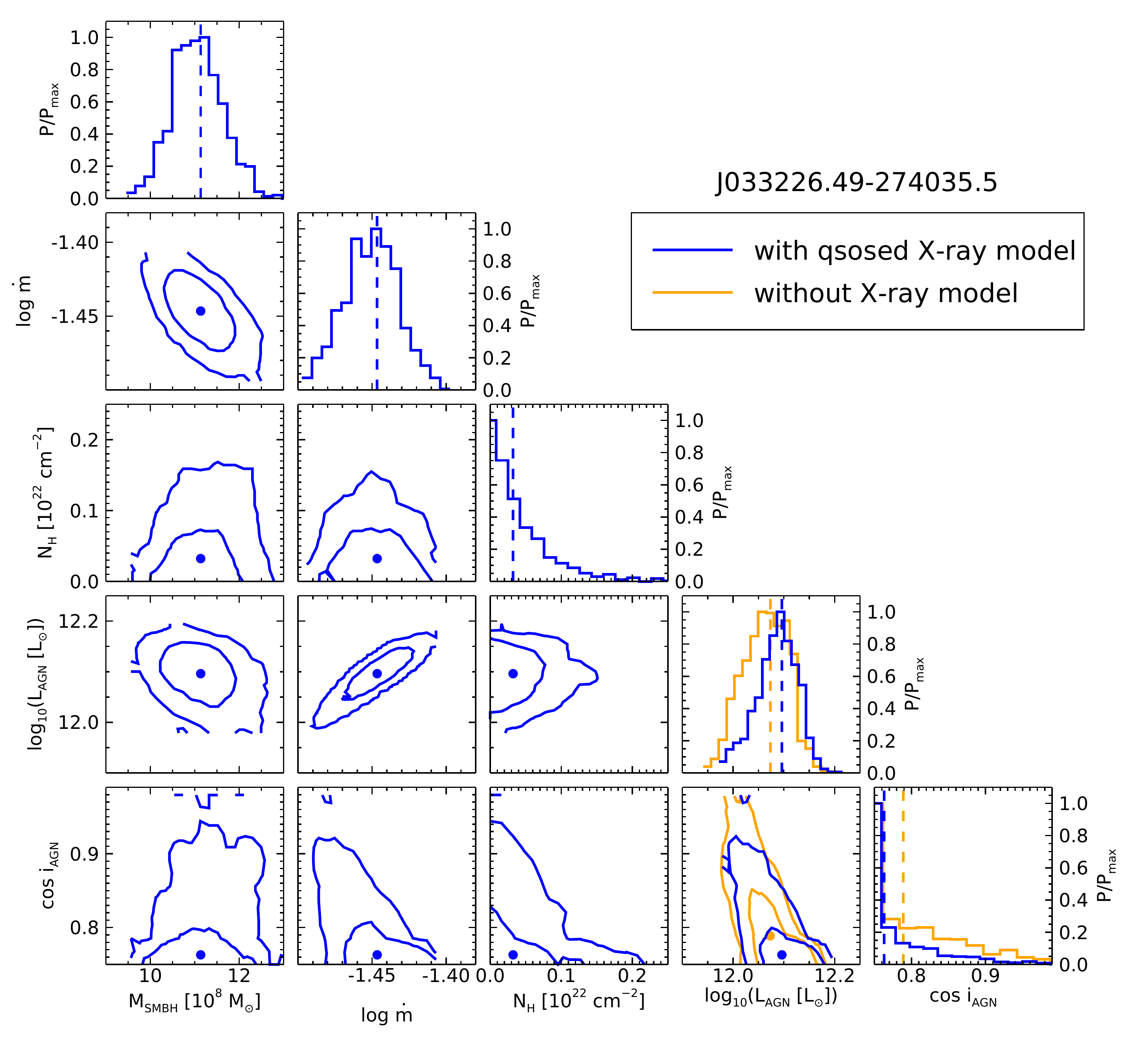}
\caption{
A corner plot of the AGN model parameters for J033226.49$-$274035.5, fit with and without an X-ray component as the blue and orange lines, respectively. The X-ray emission is modeled with the theoretical \texttt{qsosed} model and their addition to the SED results in highly consistent UV-to-IR AGN bolometric luminosity and inclination parameters compared to the fit without X-rays. In addition, the theoretical \texttt{qsosed} model also provides indirect estimates of the SMBH mass $M_{\rm SMBH}$ and Eddington ratio $\dot m = \dot M / \dot M_{\rm edd}$.
}
\label{fig:J033226_corner}
\end{figure*}

The photometry and resulting best-fitting models are shown in \autoref{fig:J033226_SEDs}. In \autoref{fig:J033226_corner}, we show a corner plot of the posterior distributions on the AGN parameters. Since the X-ray AGN model directly normalizes the UV-to-IR AGN model, we calculated the equivalent $L_{\rm AGN}$ from its UV-to-IR model to compare with the $L_{\rm AGN}$ estimated from the model without X-ray data. In the bottom right corner of the corner plot, it can be seen that the addition of the X-ray emission results in highly consistent AGN viewing angle ($\cos i_{\rm AGN}$) and luminosity ($L_{\rm AGN}$) estimates when including and excluding X-ray emission. 
Additionally, the incorporation of X-ray data also gives us estimates of $M_{\rm SMBH}$ and $\dot m$ when using the \texttt{qsosed} model, and a separate independent estimate of the obscuration of the source in $N_{\rm H}$, which indicates that this source is unabsorbed, consistent with its spectroscopic identification in the literature. In cases where MIR data is minimal or unavailable, adding the X-ray data can also give independent constraints on obscuration and add reliability to AGN IR luminosity estimates. Therefore, including X-ray observations when fitting the SED of an AGN can add valuable insights on the properties giving rise to the X-ray emission itself and place powerful constraints on the derived properties of the UV-to-IR component of the AGN.

\subsection{Stellar X-ray Emission in an Inclined Galaxy}

To demonstrate how X-rays emitted from the stellar binary population can be used to help constrain the SFR for an inclined galaxy (since X-rays are less sensitive to dust attenuation), we fit the global broadband photometry of the edge-on nearby galaxy, NGC~4631. For the UV-to-submillimeter photometry, we utilized the 30-band SINGS/KINGFISH data presented in Table~2 of \citet{2017ApJ...837...90D}. We then corrected the data for Galactic extinction before fitting, using the $E(B - V)$ values in Table~1 and $A_V$-normalized extinction values in Table~2 of \citet{2017ApJ...837...90D}.  

For the X-ray photometry, we made use of the Chandra ACIS-I data for a single $\approx$58~ks observation (ObsID~797).  These data were reduced and point-source catalogs were produced following the procedures detailed in Section~3.2 of \citet{2019ApJS..243....3L} using \texttt{CIAO} v.4.13.  To obtain spectral constraints, we utilized the \texttt{specextract} routine to extract cumulative point-source and background spectral data.  For the point-source spectrum, we chose to utilize circular apertures with radii that were four times the 90\% encircled energy fraction and centered on the 22 X-ray detected point sources within the $K_s$-band footprint of the galaxy \citep[see][for details on this region]{2003AJ....125..525J}. We extracted the background spectrum from four large regions (circles with radii spanning 1--1.5~arcmin) outside the galactic footprint that were chosen to be free of bright X-ray detected point-sources. Using \texttt{Sherpa} we fit the background-subtracted point-source spectrum using a model that consisted of both thermal and absorbed power-law components (i.e., \texttt{apec + tbabs $\times$ pow}) to account for diffuse gas and the X-ray binary emission, respectively.  Using the best-fit model, we calculated X-ray fluxes in the energy bands of 0.5--1, 1--2, 2--4, and 4--7~keV, with 10\% uncertainty on the fluxes. When fitting this X-ray data with \texttt{Lightning}, we accounted for Galactic absorption by assuming a Galactic HI column density of $N_{\rm H} = 1.29 \times 10^{20}$~cm$^{-2}$, as derived from the \texttt{prop\_colden} tool in \texttt{CIAO}.

To show the effects of including X-rays and inclination-dependence, we modeled and fit the SED using four different permutations that include or exclude X-ray data with either the \citet{2000ApJ...533..682C} or inclination-dependent attenuation curves. For all four fits, we used a stellar population with solar metallicity (i.e, $Z=0.02$) and SFH age bins of 0--10~Myr, 10--100~Myr, 0.1--1~Gyr, 1--5~Gyr, and 5--13.6~Gyr. Additionally, all fits included the \cite{2007ApJ...657..810D} dust emission model with the fixed values of $U_{\rm max}=3 \times 10^5$ and $\alpha=2$. As for the dust attenuation, which was set to be in energy balance with the dust emission, the models with the \citet{2000ApJ...533..682C} curve utilized the base curve extrapolated to the Lyman limit. For the fits with the inclination-dependent curve, we assumed the galaxy to be disk dominated and have a minimal contribution from the bulge (i.e., $B/D = 0$), a choice motivated by visual inspection. We further assume the youngest three age bins to be part of the young stellar population (i.e., $r^{\rm 0, old} = 0$ for 0~-- 1~Gyr and $r^{\rm 0, old} = 1$ otherwise). Finally, X-ray absorption was included using the \texttt{tbabs} model with \citet{2000ApJ...542..914W} abundances for the fits that included X-rays.

To fit the four models to the SED, we utilized the affine-invariant MCMC algorithm with 5\% model uncertainty, which we ran with 75 walkers for $10^4$ and $5 \times 10^4$ trials for the \citet{2000ApJ...533..682C} and inclination-dependent models, respectively. The drastic increase in trials for the inclination-dependent models is required to reach convergence (i.e., autocorrelation times $\geq 50$) of the $\cos i$ and $\tau_B^f$ parameters, since they are generally highly correlated \citep{2021ApJ...923...26D}. For all free parameters, we implemented the priors and limited initialization ranges as listed in Table~\ref{tab:NGC4631Param}. Since we know that NGC~4631 is an edge-on galaxy, we set a tabulated prior on $\cos i$ generated from the Monte Carlo method described in Section~3 of \citet{2021ApJ...923...26D}, which converts an axis ratio into a distribution of inclination. The axis ratio and its uncertainty were retrieved from HyperLeda, which provides the axis ratio calclated from the 25 mag arcsec$^{-2}$ $B$-band isophote.
\input{table_ngc4631.tex}

With the described models and algorithm, we used \texttt{Lightning} to fit each model to the SED, assuming a luminosity distance to NGC~4631 of 7.62~Mpc as given in \citet{2017ApJ...837...90D}. For each model, we set \texttt{Lightning} to automatically generate the final post-processed chain portion from the autocorrelation times and keep the final 2000 posterior samples. After confirming convergence of the fits from the autocorrelation time, we compared how the inclusion of X-rays influenced the derived properties.

\begin{figure*}[t!]
\centering
\includegraphics[width=18cm]{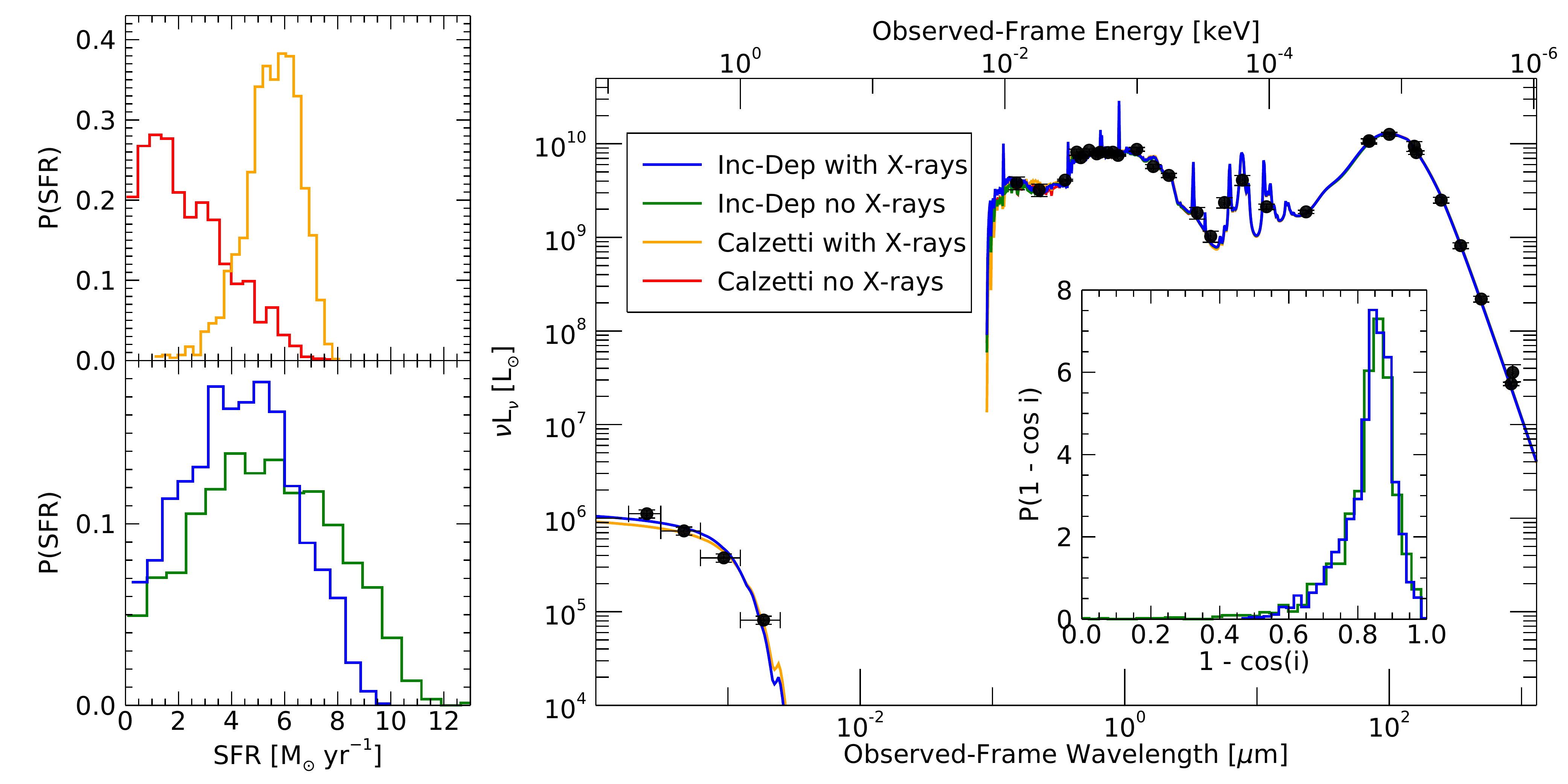}
\caption{
(\textit{Left}) Histograms of the posterior distribution functions of the SFR of the last 100 Myr for NGC~4631, which are area normalized. The \citet{2000ApJ...533..682C} models are shown in the upper panel and the inclination-dependent models are shown in the lower panel. (\textit{Right}) The total best-fit model spectrum to the observed SED for each of the four models. Inlaid in the SED plot is the posterior distributions of inclination (in terms of $1-\cos i$, where $1-\cos i = 1$ is edge-on) for the inclination-dependent models. This inlay shows that the models are correctly predicting a close to edge-on view. In all plots, the inclination-dependent model with and without X-rays and the \citet{2000ApJ...533..682C} model with and without X-rays are shown as the blue, green, orange, and red lines, respectively. While the best-fit model spectra are practically identical for each model, the resulting SFR distributions vary depending on the attenuation model and inclusion of X-ray emission.
}
\label{fig:NGC4631Plot}
\end{figure*}

In the right panels of Figure~\ref{fig:NGC4631Plot}, we show the histograms of the resulting posterior distributions of the recent SFR of the last 100 Myr. From these distributions, each of the four models can be seen to have general agreement. However, the \citet{2000ApJ...533..682C} models have a stronger variation when including the X-rays compared to the inclination-dependent models, since the \citet{2000ApJ...533..682C} attenuation model, which assumes a uniform, spherical distribution of stars and dust, is too simplistic for edge-on galaxies. Including the inclination dependence allows for a more accurate estimate of the SFR, with the inclusion of the X-rays increasing the precision of the estimate as would be expected when adding additional data. Further, the X-ray data rules out some higher SFR solutions (i.e., ${\rm SFR} > 8\ M_\odot$ yr$^{-1}$), as they become more unlikely with the X-ray data constraint.

\input{table_ngc628_components.tex}

\subsection{Comparison with other SED Fitting Codes}

\subsubsection{Bayesian Sampling Code Comparison} \label{sec:BayesCompare}

\input{table_ngc628.tex}

To show how \texttt{Lightning} compares with other Bayesian sampling SED fitting codes, we fit the global broadband photometry of the face-on nearby spiral galaxy, NGC~628 using \texttt{Lightning}, \texttt{Prospector}\footnote{\url{https://github.com/bd-j/prospector}}, and \texttt{BAGPIPES}\footnote{\url{https://github.com/ACCarnall/bagpipes}}. These two comparison codes are by no means a complete sample of the other Bayesian SED fitting codes currently available. However, we have chosen to compare directly with \texttt{Prospector} and \texttt{BAGPIPES} due to their inclusion of similar models and algorithms (e.g., non-parametric SFHs and Bayesian sampling algorithms) as \texttt{Lightning} in the interest of a ``fair'' comparison.\footnote{For a more complete comparison of the most established SED fitting codes, we recommend \citet{2022arXiv221201915P} and note that \texttt{Lightning} is not included in their comparison due to its more recent development.} While the codes can be set to have many matching components, there remain some differences between them that will generate differences in their results. We list the components of each code in Table~\ref{tab:NGC628Codes} for ease in comparison and note that the differences in IMFs are likely to cause the most variation in results \citep{2012ARA&A..50..531K, 2013ARA&A..51..393C}.

The UV-to-submillimeter photometry of NGC~628 used for this comparison was taken from the 30-band SINGS/KINGFISH data presented in Table~2 of \citet{2017ApJ...837...90D}, which we corrected for Galactic extinction using the extinction values given in Table~1 and~2 of \citet{2017ApJ...837...90D}. Since \texttt{Prospector} and \texttt{BAGPIPES} do not have a built in model uncertainty method like \texttt{Lightning}, we added in quadrature an additional 10\% uncertainty to the quoted uncertainties in \citet{2017ApJ...837...90D} to act as model uncertainties in our fits. We note that this is the typically utilized method in SED fitting to account for model uncertainties, and therefore it is a reasonable method for accounting for additional uncertainty not contained within the data.

For the models in each code, we used a stellar population, as given in Table~\ref{tab:NGC628Codes}, with constant solar metallicity (i.e, $Z=0.02$) and SFH age bins of 0--10~Myr, 10--100~Myr, 0.1--1~Gyr, 1--5~Gyr, and 5--13.4~Gyr. We attenuated the stellar emission in all codes using the original \citet{2000ApJ...533..682C} curve extrapolated to the Lyman limit. Finally, the dust attenuation was set to be in energy balance with the \cite{2007ApJ...657..810D} dust emission model, where $\alpha=2$, $U_{\rm max}=3 \times 10^5$ for \texttt{Lightning}, and $U_{\rm max}=10^6$ for \texttt{Prospector} and \texttt{BAGPIPES} (the differences in $U_{\rm max}$ have minimal effects on the models, see Section~\ref{sec:DustEmission}).

To fit the SED with each code, we utilized their Bayesian sampling algorithms to generate posterior distributions containing 2000 samples. For \texttt{Lightning} and \texttt{Prospector}, we ran their affine-invariant MCMC algorithm for $10^4$ trials with 75 walkers, which was sufficient for each code to reach convergence using the autocorrelation times. The resulting chains were then post-processed (i.e., burn-in removed and thinned) to the 2000 samples using the longest autocorrelation time of any parameter.\footnote{\texttt{Prospector} also has a nested sampling algorithm. However, we chose to use the affine-invariant MCMC algorithm to have the closest possible comparison with \texttt{Lightning}.} For \texttt{BAGPIPES}, we used the available \texttt{MultiNest} \citep{2009MNRAS.398.1601F, 2019OJAp....2E..10F} nested sampling algorithm, which we ran using 1000 live points. We note that we increased the number of live points from the default of 400 to 1000 after testing showed that we could get significant variation in the posteriors between runs when using fewer than 1000 live points for our chosen model. In Table~\ref{tab:NGC628Param}, we list the free parameters in each model along with their utilized prior function. We note that we used the same priors across each code, except for the SFH parameters. Unlike \texttt{Lightning}, \texttt{Prospector} and \texttt{BAGPIPES} normalize their SFHs by stellar mass rather than SFR. Therefore, they require different priors to accommodate the change in normalization, which is expected to cause minimal to no effects on the fits due to the utilization of the same SFH bins.

With the results from the fits, we first compared the computational performance of each code. All codes were run sequentially on the author's 2016 MacBook, which contains a 2-core, 1.2 GHz CPU. \texttt{Lightning}, \texttt{Prospector}, and \texttt{BAGPIPES} took 1279.7~s, 4864.2~s, and 920.0~s, respectively, to complete their fitting. While \texttt{BAGPIPES} can be seen to be almost 1.4 times faster than \texttt{Lightning} (which is 3.8 times faster than \texttt{Prospector}), it is important to note that this is due to the fewer likelihood evaluations needed by the nested sampling algorithm, which was designed (in part) to reduce the number of model evaluations required to produce a full sampling of the posterior \citep[see, e.g.,][]{2009MNRAS.398.1601F, 2019OJAp....2E..10F}. Where \texttt{Lightning} and \texttt{Prospector} each took $1.125 \times 10^6$ likelihood evaluations, \texttt{BAGPIPES} only performed an order of magnitude fewer (115,921 evaluations) to fit. Therefore, the difference in algorithms allowed for an overall similar fitting time as \texttt{Lightning}. However, in terms of likelihood evaluations per second (which is a better comparison of the practical speed of different SED fitting codes), \texttt{Lightning} is almost seven times faster than \texttt{BAGPIPES}, which is a result of its designed computational efficiency.
\begin{figure}[t!]
\centering
\includegraphics[width=8.5cm]{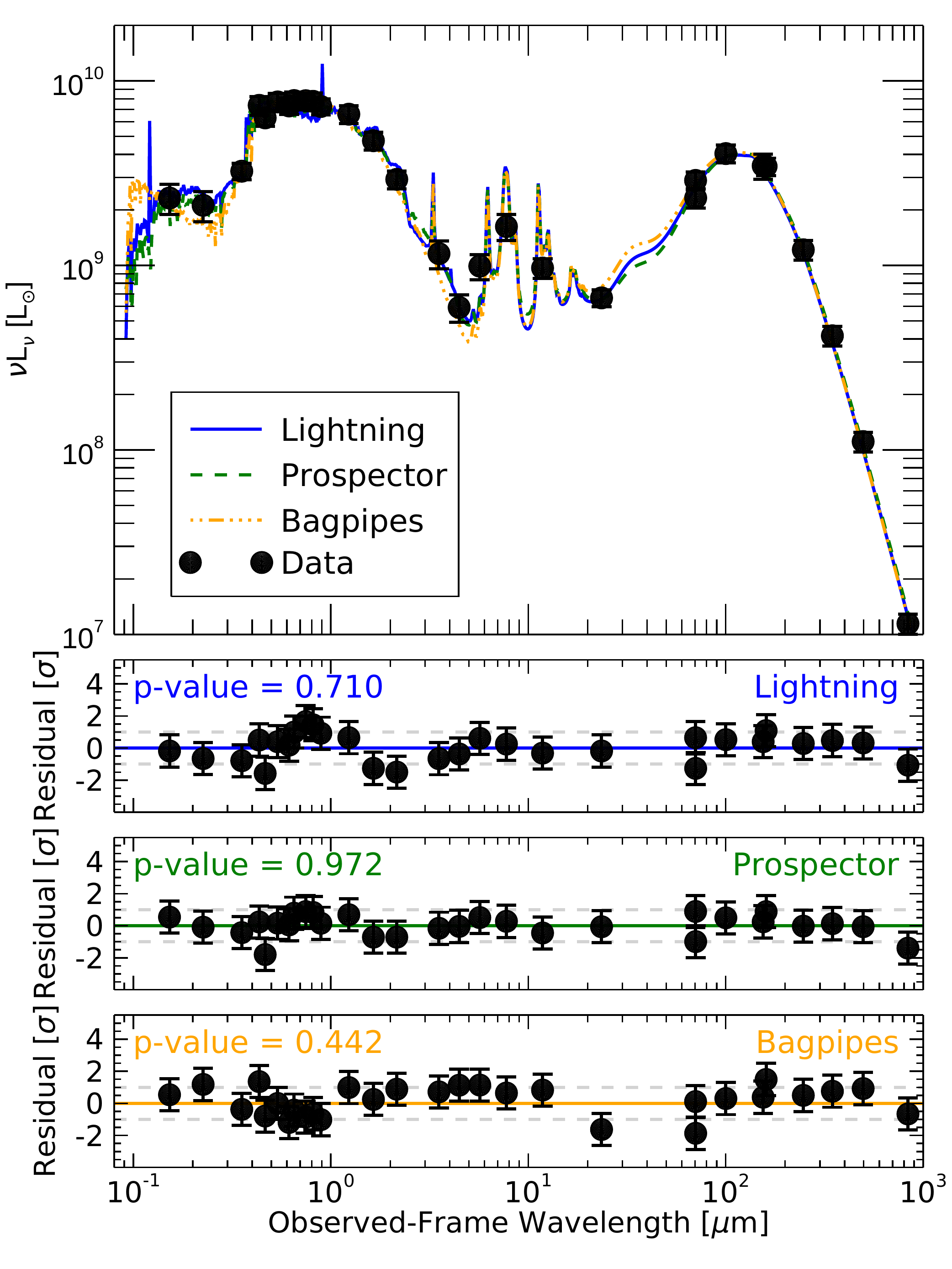}
\caption{
The total best-fit model spectra to the observed SED of NGC~628 and the associated residuals as generated by \texttt{Lightning}, \texttt{Prospector}, and \texttt{BAGPIPES} as the blue, green, and orange lines, respectively. For each code, the $p$-value from a PPC is shown in the upper left of each residual plot. In general, it can be seen that all three codes model the data well, both from the PPC and the residuals. 
}
\label{fig:NGC0628SEDs}
\end{figure}

To show how the derived results compare at a base level, we show the observed photometry and best-fitting model spectra in Figure~\ref{fig:NGC0628SEDs}. In the lower residual panels for each fit, we also show the derived $p$-value for the fits calculated from a PPC. From the $p$-values and residuals, it can be seen that all codes appropriately model the data. It is interesting to note that the best-fit model spectra and resulting residuals from the \texttt{Lightning} and \texttt{Prospector} fits are highly similar, since their models only differ by the SSPs. However, \texttt{BAGPIPES} identifies a comparatively unique best-fit spectrum and residuals in the UV-to-NIR. This variation is expected, since the different IMF in \texttt{BAGPIPES} can create a substantial variation in the UV-to-NIR stellar emission models.
\begin{figure*}[t!]
\centering
\includegraphics[width=18cm]{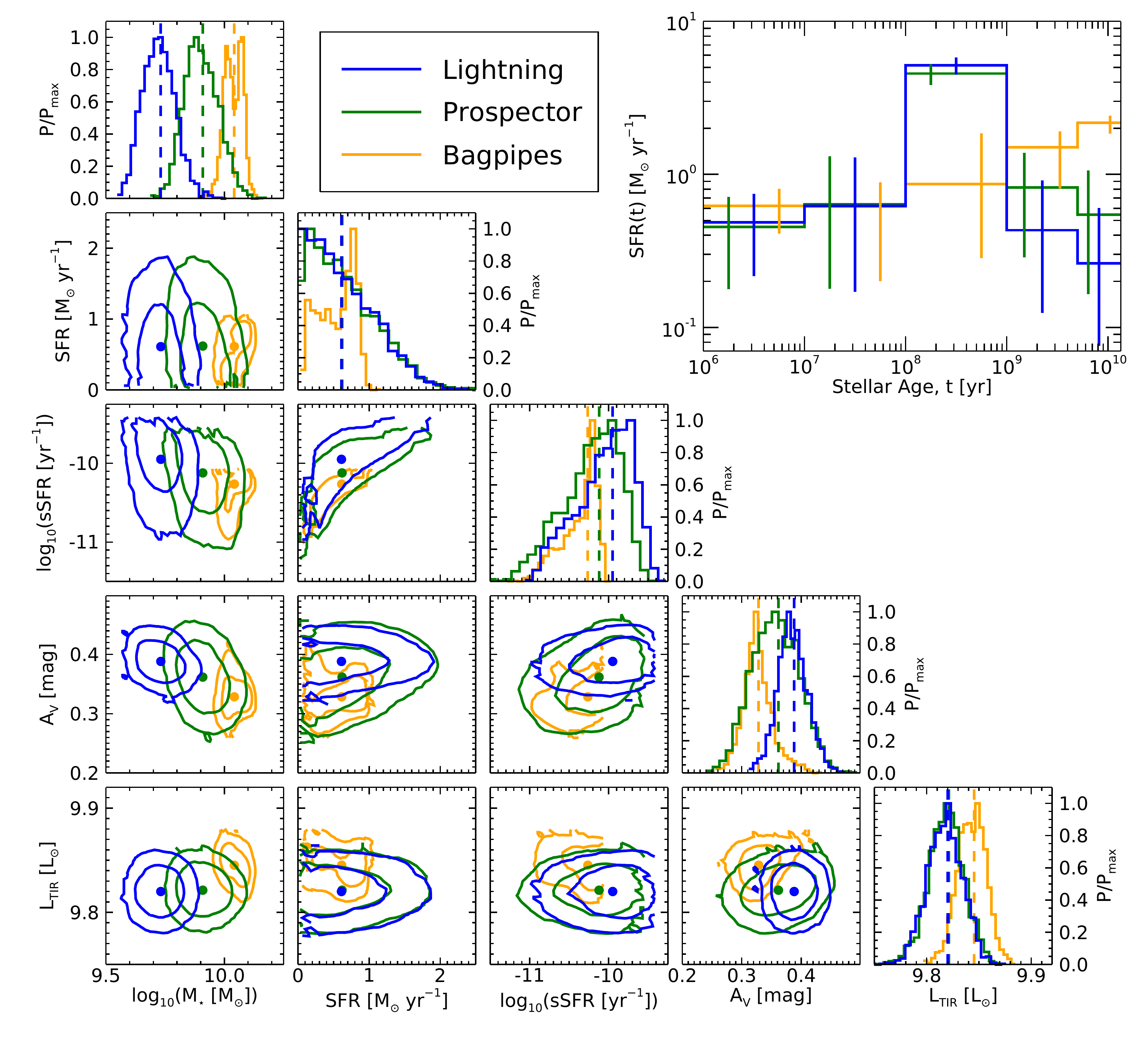}
\caption{
(\textit{Lower Left}) A corner plot of the derived parameters for NGC~628. These parameters include the stellar mass ($M_\star$), the recent SFR of the last 100 Myr, the sSFR of the last 100 Myr, the $V$-band attenuation ($A_V$), and the $L_{\rm TIR}$ for each SED fitting code. (\textit{Upper Right}) The median SFH with the 16th and 84th percentile uncertainty range given as the offset vertical lines for each SED fitting code. For all plots, the results from \texttt{Lightning}, \texttt{Prospector}, and \texttt{BAGPIPES} are given as the blue, green, and orange lines, respectively. From these plots, it can be seen that the results from \texttt{Lightning} and \texttt{Prospector} are highly consistent, while the \texttt{BAGPIPES} results vary in consistency, especially for the SFH.
}
\label{fig:NGC0628SFH}
\end{figure*}

\input{table_CIGALEparams.tex}

Finally, in Figure~\ref{fig:NGC0628SFH}, we show the derived SFHs and posterior distributions for five commonly derived parameters from SED fitting: the (surviving) stellar mass ($M_\star$), the SFR of the last 100~Myr, the sSFR of the last 100~Myr, the $V$-band attenuation ($A_V$), and $L_{\rm TIR}$ (i.e., the bolometric luminosity from 8--1000~$\mu$m). From these distributions in the lower left, it can be seen that all three codes have derived parameters, except stellar mass, that are in excellent agreement. The variation in stellar mass between codes is expected and entirely due to the differences in SSPs and IMFs, which dictate the surviving stellar mass of the populations over time. As for the other parameters and SFHs, \texttt{Lightning} and \texttt{Prospector} have near identical results, which is a quality indicator given that both codes were run using almost identical models and were independently developed. As for the differences with \texttt{BAGPIPES}, these are a combination of the differences in fitting algorithms and IMFs, especially for the SFH. However, overall for both \texttt{Lightning} and SED fitting as a whole, it is reassuring that codes with different models and algorithms generally return derived parameters that are in statistical agreement.

\subsubsection{X-ray AGN Code Comparison}

With the new additions to \texttt{Lightning} making it capable of fitting the X-ray-to-IR SEDs of AGNs, we examined how its derived AGN properties compared to the only other SED fitting code publicly available (at the time of publication)\footnote{Other X-ray to IR AGN SED-fitting codes exist \citep[e.g. \texttt{ARXSED}][]{2023ApJ...945..145A}, but they are not publicly available.} also capable of modeling X-ray-to-IR AGN emission, \texttt{CIGALE}\footnote{\url{https://cigale.lam.fr}} \citep{2019A&A...622A.103B, 2020MNRAS.491..740Y, 2022ApJ...927..192Y}. Unlike the comparison in Section~\ref{sec:BayesCompare}, \texttt{Lightning} and \texttt{CIGALE} have notably different models and algorithms. For example, commonalities for both include the modified \citet{2000ApJ...533..682C} attenuation curve, \citet{2007ApJ...657..810D} dust emission models, SKIRTOR AGN templates, and power-law X-ray AGN models. However, \texttt{Lightning} implements non-parametric SFHs, while \texttt{CIGALE} mainly relies on parametric SFHs. Additionally, when fitting the models to data, \texttt{Lightning} can utilize either Bayesian sampling or maximum-likelihood inferencing, while \texttt{CIGALE} only utilizes a gridded Bayesian statistical inferencing method. Therefore, rather than attempting to use similar models and algorithms like we did in Section~\ref{sec:BayesCompare}, in this example, we fit using the recommended settings of each code when modeling sources that include an X-ray AGN to see how the derived results compare.

For this comparison, we utilized the same X-ray AGN presented in Section~\ref{sec:DeepAGN}, J033226.49$-$274035.5. The fit we performed in Section~\ref{sec:DeepAGN} already utilized our recommended models and the vast majority of our recommended model settings (see Table~\ref{tab:J033226Param}). The only difference from the recommendation is fixing $q_{\rm PAH}$ and limiting the prior range of $\cos i_{\rm AGN}$. We retain these setting for this example, as $q_{\rm PAH}$ would be a nuisance parameter if not fixed (as can be seen from the negligible contribution to the MIR by the dust emission in Figure~\ref{fig:J033226_SEDs}) and the $\cos i_{\rm AGN}$ prior restrics modeling to type~1 AGN. Therefore, we utilized our results from the X-ray model fit in Section~\ref{sec:DeepAGN} for this comparison.

For the \texttt{CIGALE} settings, we utilized their recommended settings for an X-ray AGN using the example in their online documentation\footnote{\url{https://gitlab.lam.fr/cigale/manual/-/blob/master/examples/akari_nep_xray_agn/pcigale.ini}}. This model consisted of a delayed exponential SFH with a \citet{2003PASP..115..763C} IMF and metallicity of $Z=0.02$, UV-to-IR AGN emission using the SKIRTOR templates with the \citet{2005A&A...437..861S} intrinsic-disk type, X-ray AGN emission from a power-law model assuming $\Gamma = 1.8$, \citet{2000ApJ...533..682C} dust attenuation, and \citet{2014ApJ...784...83D} dust emission. The list of all gridded or non-default parameters used for each model are given in Table~\ref{tab:CIGALE}. We note that we deviated slightly from their recommended settings for two parameters in the model: $\theta$ (the AGN viewing angle, equivalent to $i_{\rm AGN}$ in \texttt{Lightning}) and $\delta_{\rm AGN}$ (the deviation from the default UV/optical slope of the AGN model). We limit the grid points for $\theta$ to all possible options for type~1 AGN views and allow $\delta_{\rm AGN}$ to vary as is recommended when modeling type~1 AGN \citep{2022ApJ...927..192Y}. Additionally, we set ``lambda\_fracAGN'' to define the AGN IR fraction as the integrated luminosity between observed-frame 5--1000~$\mu$m for ease in comparison with \texttt{Lightning}.

Utilizing the same observational data of J033226.49$-$274035.5 given in Section~\ref{sec:DeepAGN}, we then fit the  data using \texttt{CIGALE}. We do note that while we used the same UV-to-IR data, \texttt{CIGALE} requires X-ray data be input as absorption-corrected X-ray fluxes rather than the instrumental counts input into \texttt{Lightning}. This conversion to absorption-corrected fluxes requires one to either assume or fit a spectral model to the X-ray spectrum prior to performing the full SED fit, in order to include the X-ray data. For this example, we chose to fit the X-ray spectrum with an absorbed power-law (i.e. $\tt tbabs \times pow$) using \textsc{Sherpa} v4.13, finding a best-fit $\Gamma = 1.96$ and $N_{\rm H} < 1 \times 10^{20}\ {\rm cm^{-2}}$. From this X-ray spectrum fit, we then calculated the absorption-corrected fluxes for input into \texttt{CIGALE} in 15 log-spaced bands from 0.5--6.0 keV (the same energy bands used in Section~\ref{sec:DeepAGN}).

Both codes were then run sequentially on the author's 2016 MacBook, which contains a 2-core, 1.2 GHz CPU. \texttt{Lightning} and \texttt{CIGALE} took $\approx$1~hr and $\approx$10~hr, respectively, to complete their fitting. While \texttt{Lightning} fit an order of magnitude faster in absolute terms, like in Section~\ref{sec:BayesCompare}, it is better to compare the likelihood evaluations per time interval for an equivalent speed comparison. For this fit, \texttt{Lightning} performed $3 \times 10^6$ likelihood evaluations, while \texttt{CIGALE} performed a factor of four more (12,830,400 evaluations) to fit. Therefore, \texttt{Lightning} is approximately two times faster in terms of likelihood evaluations, which again is a result of its designed computational efficiency.

\begin{figure}[t!]
\centering
\includegraphics[width=8.5cm]{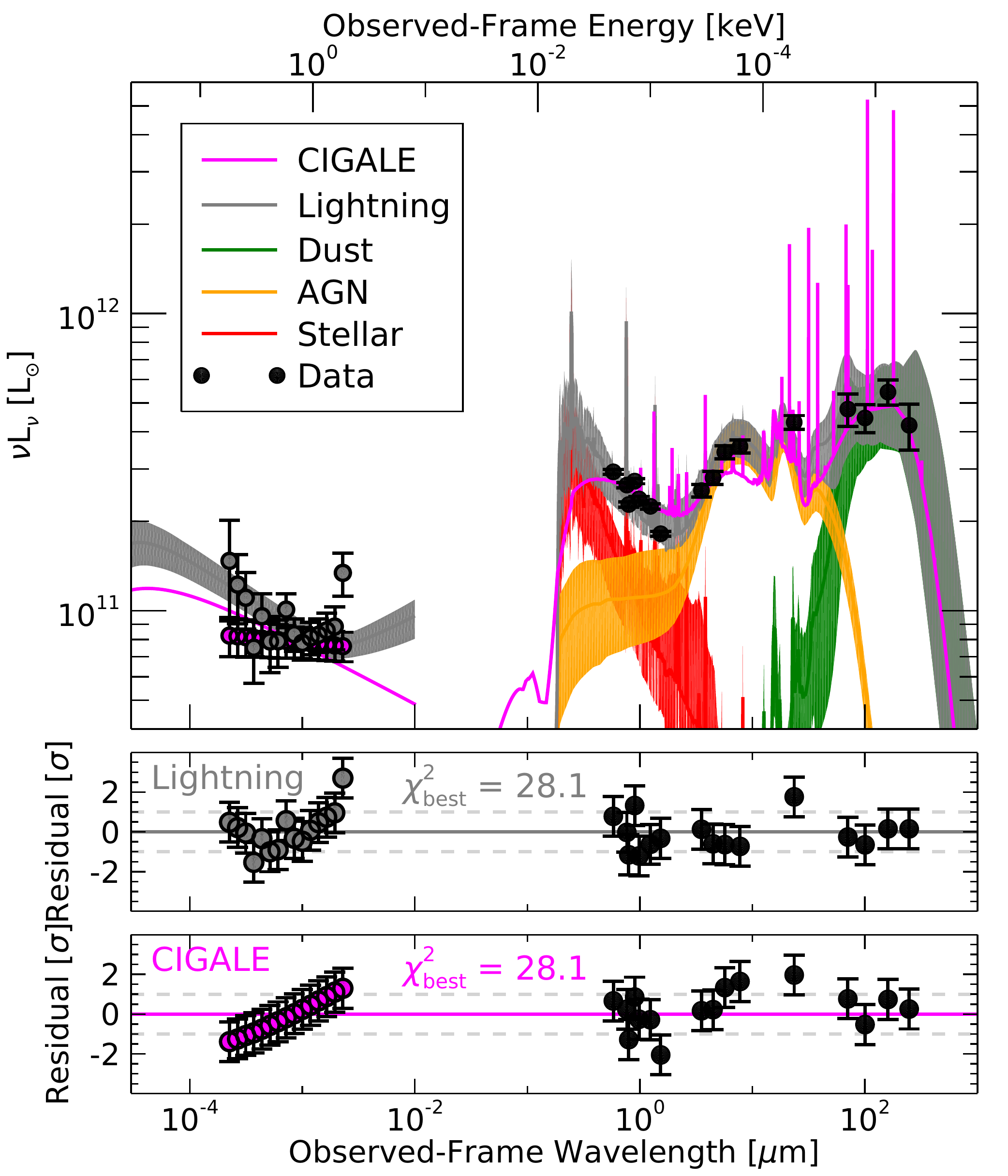}
\caption{
The total best-fit model spectra to the observed SED of J033226.49$-$274035.5 and the associated residuals as generated by \texttt{Lightning} and \texttt{CIGALE} as the gray and magenta lines, respectively. The gray and magenta colored data points in the X-rays correspond to each codes input data, since \texttt{Lightning} had inputs in units of observed counts (which are converted to model dependent luminosities for plotting) and \texttt{CIGALE} had inputs in units of absorption-corrected fluxes. For the \texttt{Lightning} fit, we include each model component as in Figure~\ref{fig:J033226_SEDs}. Additionally, we include the range of the top 64\% of best-fit models from the MCMC posterior as the correspondingly colored shaded regions to show the uncertainty range on the component SEDs. For each code, we also show the best-fit $\chi^2$ value in the center of each residual plot. In general, it can be seen that both codes model their data well from the best-fit $\chi^2$ and the residuals. 
}
\label{fig:LvCSED}
\end{figure}

\input{table_Lightnging_v_CIGALE.tex}

With the results from both fits, we compared the derived physical properties from \texttt{Lightning} to those derived by \texttt{CIGALE}. In Figure~\ref{fig:LvCSED}, we show the best-fit model SED for both codes, with the gray and magenta colored data points in the X-rays correspond to each codes input data, since \texttt{Lightning} had inputs in units of observed counts (which are converted to model dependent luminosities for plotting) and \texttt{CIGALE} had inputs in units of absorption-corrected fluxes. Additionally, since \texttt{Lightning} uses an MCMC algorithm, we generated the range of the top 64\% of the best-fit models, in terms of the posterior, to create an uncertainty range on the model SEDs. With the uncertainty range, it can be seen that the best-fit model SED for both codes are highly consistent in the UV-to-IR, but differ in the X-rays. This difference in the X-ray is mainly caused by the difference in models (\texttt{Lightning} uses the more flexible \texttt{qsosed} model, while \texttt{CIGALE} uses a more rigid power-law) and input data (\texttt{Lightning} uses the observed counts, while \texttt{CIGALE} uses absorption-corrected fluxes). Therefore, differences are to be expected for the models in the X-rays.

Finally, we list the commonly quoted parameters (mainly focusing on X-ray and AGN parameters) derived from each code in Table~\ref{tab:LightvCIGALE}. These parameters include: the recent SFR over the last 100 Myr, the surviving stellar mass ($M_\star$), the total integrated IR luminosity of the dust emission model ($L_{\rm TIR}$), the AGN torus inclination ($i_{\rm AGN}$), the UV-to-IR AGN bolometric luminosity ($L_{\rm AGN}$), the fraction of AGN IR luminosity to the total IR luminosity from 5--1000~$\mu$m ($\rm frac_{AGN}$), and the slope of the $\tilde{L}_{2\rm\ keV}-\tilde{L}_{2500}$ relationship ($\alpha_{\rm OX}$). Since \texttt{Lightning} does not calculate $\rm frac_{AGN}$ or $\alpha_{\rm OX}$ by default, we derived these values from the postprocessed results. We calculated $\rm frac_{AGN}$ as the fraction of the model AGN IR luminosity to the total IR luminosity between observed-frame 5--1000~$\mu$m. For $\alpha_{\rm OX}$, we first derived monochromatic luminosities at rest-frame 2~keV and 2500~\AA and computed it as $\alpha_{\rm OX} = -0.3838 \log_{10}(L_{2500{\rm \AA}}/L_{2{\rm keV}})$ (Equation~5 in \citealt{2020MNRAS.491..740Y}).

From the data in Table~\ref{tab:LightvCIGALE}, it can be seen that these commonly used properties are in excellent statistical agreement between the two codes. Of the derived properties, the SFRs and stellar masses have the largest uncertainty. This is caused by the uncertainty in the stellar models, since the type~1 AGN contributes a significant fraction of the emission at optical wavelengths. As for the AGN parameters, we find that only $L_{\rm AGN}$ disagrees between codes at $\geq$1$\sigma$ level, but their values are similar. The slight disagreement arises from the small uncertainties for both fits that are caused by the highly constraining 15 band X-ray data, which strongly limits the allowable AGN luminosity in the UV-to-IR. Comparatively, $\rm frac_{AGN}$ is highly consistent between codes, indicating common contributions from the AGN to the total IR model. Overall, just like the comparison in Section~\ref{sec:BayesCompare}, it is reassuring, for these results and SED fitting in general, that both \texttt{Lightning} and \texttt{CIGALE} return derived parameters that are in excellent statistical agreement despite using different models and algorithms.

\section{Summary and Planned Additions} \label{sec:Summary}

In this paper, we have presented the most recent version of the SED fitting code \texttt{Lightning}. The new version of \texttt{Lightning} contains a variety of models and algorithms that can be used to account for any combination of stellar, dust, and AGN emission in an observed X-ray to submillimeter SED. A brief review of each of these models and algorithms is as follows:

\begin{enumerate}
\item Stellar emission can be modeled using the SSPs from \texttt{P\'EGASE} integrated over the age bins given by the user-defined non-parametric SFH. Stellar X-ray emission from the XRBs is linked to the SFH using a power-law spectral model and the empirical parameterizations of $L_X / M_\star$ with stellar age given in \citet{2022ApJ...926...28G}.

\item AGN emission can be modeled in the UV-to-IR using a subset of the SKIRTOR models. X-ray AGN emission can be modeled as (1) a simple rigid power-law spectra, which is tied to the UV-to-IR AGN emission using the \citet{2017A&A...602A..79L} $\tilde{L}_{2\rm\ keV}-\tilde{L}_{2500}$ relationship or (2) the physically-motivated \texttt{qsosed} models from \citet{2018MNRAS.480.1247K}, which directly scale the UV-to-IR emission as a function of the mass and Eddington ratio of the SMBH.

\item Dust attenuation of the UV-to-NIR emission can be modeled using either a modified form of the \citet{2000ApJ...533..682C} curve or the inclination-dependent curve described in \citet{2021ApJ...923...26D}. Absorption of X-ray emission, when included, is modeled using either the \texttt{tbabs} or the \texttt{Sherpa atten} models.

\item Dust emission can be modeled using the \citet{2007ApJ...657..810D} model. When included, dust emission can be set to be in energy balance with the dust attenuation, which requires the bolometric luminosity of the dust emission to be equal to the bolometric luminosity of the attenuated light. 

\item Algorithms for fitting the models to the data include both maximum likelihood and Bayesian methods. For the maximum likelihood method, \texttt{Lightning} uses the \texttt{MPFIT} implementation of the gradient-descent Levenberg–Marquardt algorithm. For the Bayesian methods, \texttt{Lightning} includes two MCMC algorithms, an adaptive Metropolis-Hastings algorithm from \citet{Andrieu2008} and an implementation of the \citet{2010CAMCS...5...65G} affine-invariant algorithm.
\end{enumerate}

With these models and algorithms, we presented different example applications of \texttt{Lightning}. These examples included (1) deriving spatially resolved stellar properties of M81 using an SED map, (2) demonstrating how the SMBH properties of an AGN can be derived by including X-ray emission, (3) exploring how X-ray emission and inclination-dependent attenuation can be used to constrain the SFR of an edge-on galaxy, (4) comparing the performance of \texttt{Lightning} to similar Bayesian sampling SED fitting codes (\texttt{Prospector} and \texttt{BAGPIPES}), and (5) comparing the X-ray-to-IR AGN properties derived from \texttt{Lightning} and \texttt{CIGALE}. From these examples, we clearly demonstrate the capabilities of \texttt{Lightning} and some of its potential uses.

In future updates to \texttt{Lightning}, we plan to expand our current \texttt{P\'EGASE} stellar models. This will include adding additional IMF choices and allowing for a constant but continuous metallicity, which, like other SED fitting codes, we plan to have as an optional free parameter. Additionally, \texttt{Lightning} is currently restricted to using an exact redshift value if it is used as the distance indicator. We plan to allow for increased flexibility in redshift, where it can have an associated prior distribution when fitting using a Bayesian method. This would allow for better propagation of uncertainty when using photometric redshifts, which can have large associated uncertainties.

Further, since \texttt{Lightning} was originally developed to be used in XRB population studies, we plan to add new SSPs like \texttt{BPASS}\footnote{\url{https://bpass.auckland.ac.nz}} \citep{2017PASA...34...58E, 2018MNRAS.479...75S} and/or \texttt{POSYDON}\footnote{\url{https://posydon.org}} \citep{2022arXiv220205892F} that include binary population evolution. This would allow for more accurate stellar emission models, as binary populations can significantly influence the stellar emission \citep{2020arXiv200511883E}. Additionally, binary stars are the progenitors of compact object binaries. Future versions of binary stellar population models may provide predictions for the observed X-ray binary luminosity function and its evolution with age and metallicity, based on the same prescriptions that govern the age and metallicity evolution of the stellar population. By adopting such models, we can self-consistently produce $L_X / M_{\star}$ with our stellar population models rather than relying on empirical relations.

\begin{acknowledgements}
We acknowledge and thank the anonymous referee for their valuable comments that helped improve the quality of this paper. We gratefully acknowledge support from the NASA Astrophysics Data Analysis Program (ADAP) grant 80NSSC20K0444 (KD, EBM, RTE, BDL). KG is supported by an appointment to the NASA Postdoctoral Program at NASA Goddard Space Flight Center, administered by Oak Ridge Associated Universities under contract with NASA. ABZ supported by NASA award number 80GSFC21M0002. This work has made use of the Arkansas High Performance Computing Center, which is funded through multiple National Science Foundation grants and the Arkansas Economic Development Commission. We acknowledge the usage of the HyperLeda database (\url{http://leda.univ-lyon1.fr}).
\end{acknowledgements}

\software{\texttt{BAGPIPES} \citep{2018MNRAS.480.4379C}; \texttt{CIAO} \citep{2006SPIE.6270E..1VF}; \texttt{Lightning} \citetext{\citealp{2017ApJ...851...10E, 2021ApJ...923...26D}; Monson et al. 2023, submitted}; \texttt{MPFIT} \citep{1978LNM...630..105M, 2009ASPC..411..251M}; \texttt{P\'EGASE} \citep{1997A&A...326..950F}; \texttt{Prospector} \citep{2021ApJS..254...22J}; \texttt{Sherpa} \citep{sherpa4.13.0}}

\bibliographystyle{aasjournal}

\bibliography{Lightning}

\end{document}

%% file: table_feature.tex
\begin{deluxetable}{lr}
\tabletypesize{\footnotesize}
\tablecolumns{2}
\tablecaption{\label{tab:features}References for Previously Published Articles Describing the Implementation of Individual \texttt{Lightning} Features.}
\tablehead{\colhead{Feature} & \colhead{Reference}}
\startdata
\cutinhead{Models}
Simple Stellar Populations & \citet{2017ApJ...851...10E} \\
Non-parametric SFH & \citet{2017ApJ...851...10E} \\
Stellar X-ray Emission & \citet{2023ApJ...951...15M} \\
UV-to-IR AGN Emission & \citet{2023ApJ...951...15M} \\
X-ray AGN Emission & \citet{2023ApJ...951...15M} \\
\citet{2000ApJ...533..682C} Attenuation & \citet{2017ApJ...851...10E} \\
Inclination-dependent Attenuation & \citet{2021ApJ...923...26D} \\
X-ray Absorption & \citet{2023ApJ...951...15M} \\
\citet{2007ApJ...657..810D} Dust Emission & \citet{2021ApJ...923...26D} \\
\cutinhead{Fitting Algorithms}
Gradient Descent (\texttt{MPFIT}) & This work \\
Adaptive MCMC & \citet{2021ApJ...923...26D} \\
Affine-Invariant MCMC & \citet{2023ApJ...951...15M} \\
\enddata
\end{deluxetable}

%% file: table_SEDparams.tex
\begin{deluxetable*}{lcccl}[t!]
\tabletypesize{\footnotesize}
\tablecolumns{5}
\tablecaption{\label{tab:SEDParams}Summary of Possible Free Parameters for Each Model SED Component in \texttt{Lightning}.}
\tablehead{\colhead{Physical Model Component} & \colhead{Parameter} & \colhead{Units\tablenotemark{a}} & \colhead{Allowed Range} & \colhead{Description}}
\startdata
Stellar Emis.                          & $\psi_j$                & $M_\odot$ yr$^{-1}$ & $[0, \infty)$           & Star formation history coefficients \\
\citet{2000ApJ...533..682C} Atten.   & $\tau_{{\rm DIFF}, V}$  & \nodata              & $[0, \infty)$           & $V$-band optical depth of diffuse dust \\
                                          & $\delta$                & \nodata              & $(-\infty, \infty)$     & Power-law slope deviation \\
                                          & $\tau_{{\rm BC}, V}$    & \nodata              & $[0, \infty)$           & $V$-band optical depth of the birth cloud \\
SKIRTOR UV-to-IR AGN Emis.\tablenotemark{b}             & $\log_{10} L_{\rm AGN}$\tablenotemark{c} & $\log_{10} L_\odot$ & $[0, 20]$               & UV-to-IR integrated AGN luminosity \\ 
                                          & $\tau_{9.7}$            & \nodata             & $[3, 11]$               & Edge-on optical depth of the AGN torus at 9.7 $\mu$m \\
                                          & $\cos i_{\rm AGN}$      & \nodata             & $[0, 1]$                & Cosine of the line-of-sight AGN torus inclination  \\
\texttt{qsosed} X-ray AGN Emis.        & $M_{\rm SMBH}$          & $M_\odot$           & $[10^5, 10^{10}]$       & SMBH mass \\
                                          & $\log_{10} \dot m$      & \nodata             & $[-1.5, 0.3]$           & SMBH accretion rate, Eddington rate normalized  \\
Inclination-dependent Atten.\tablenotemark{b}   & $\cos i$     & \nodata              & $[0, 1]$                & Cosine of the line-of-sight galaxy inclination \\
                                          & $\tau_B^f$              & \nodata              & $[0, 8]$                & Central $B$-band face-on optical depth \\
                                          & $B/D$                   & \nodata              & $[0, \infty)$           & Bulge-to-disk ratio \\
                                          & $F$                     & \nodata              & $[0, 0.61]$             & Clumpiness factor \\
X-ray Absorption                          & $N_{\rm H}$                   & $10^{20}$ cm$^{-2}$ & $[10^{-4}, 10^5]$       & HI column density along the line of sight \\
\citet{2007ApJ...657..810D} Dust Emis. & $\alpha$                &  \nodata             & $[-10, 4]$              & Power law slope of the intensity distribution \\
                                          & $U_{\rm min}$           &  \nodata             & $[0.1, 25]$             & Minimum radiation field intensity \\
                                          & $U_{\rm max}$           &  \nodata             & $[10^3, 3 \times 10^5]$ & Maximum radiation field intensity \\
                                          & $\gamma$                &  \nodata             & $[0, 1]$                & Dust mass fraction illuminated from $U_{\rm min}$ to $U_{\rm max}$ \\
                                          & $q_{\rm PAH}$           &  \nodata             & $[0.0047, 0.0458]$      & Mass fraction of PAHs in dust mixture \\
                                          & $L_{\rm TIR}$\tablenotemark{d}           & $L_\odot$           & $[0, \infty)$           & Total integrated IR luminosity \\
\enddata
\tablenotetext{a}{Parameters without units are unitless.}
\tablenotetext{b}{The inclination-dependent attenuation and SKIRTOR AGN models are currently not compatible.}
\tablenotetext{c}{$L_{\rm AGN}$ is only a free parameter if fitting without the \texttt{qsosed} X-ray AGN model.}
\tablenotetext{d}{$L_{\rm TIR}$ is only a free parameter if fitting without energy balance.}
\end{deluxetable*}

%% file: table_m81.tex
\begin{deluxetable}{lcc}[t]
\tabletypesize{\footnotesize}
\tablecolumns{3}
\tablecaption{\label{tab:M81Param}Summary of Parameters Used in the M81 Example.}
\tablehead{\colhead{Parameter} & \colhead{Prior Function} & \colhead{Initialization Range}}
\startdata
$\psi_j$                & $\mathcal{U}(0, 10^3)$        & $[0, 1]$           \\
$\tau_{{\rm DIFF}, V}$  & $\mathcal{U}(0, 10)$          & $[0, 1]$           \\
$\delta$                & $\mathcal{U}(-2.3, 0.4)$        & $[-1, 0]$          \\
$\tau_{{\rm BC}, V}$    & Fixed                         & $0$                \\
$\alpha$                & Fixed                         & $2$                \\
$U_{\rm min}$           & $\mathcal{U}(0.1, 25)$        & $[0.1, 5]$         \\
$U_{\rm max}$           & Fixed                         & $3 \times 10^5$    \\
$\gamma$                & $\mathcal{U}(0, 1)$           & $[0, 0.1]$         \\
$q_{\rm PAH}$           & $\mathcal{U}(0.0047, 0.0458)$ & $[0.0047, 0.0458]$ \\
\enddata
\tablecomments{$\mathcal{U}(a, b)$ indicates a uniform distribution from $a$ to $b$. Fixed parameters have their value listed in the
initialization range column.}
\end{deluxetable}

%% file: table_J033226.tex
\begin{deluxetable}{lcc}[t!]
\tabletypesize{\footnotesize}
\tablecolumns{3}
\tablecaption{\label{tab:J033226Param}Summary of Parameters Used in the J033226.49$-$274035.5 Example.}
\tablehead{\colhead{Parameter} & \colhead{Prior Function} & \colhead{Initialization Range}}
\startdata
\cutinhead{All Models}
$\psi_j$                & $\mathcal{U}(0, 10^3)$         & $[0, 100]$         \\
$\tau_{{\rm DIFF}, V}$  & $\mathcal{U}(0, 10)$           & $[0, 3]$           \\
$\alpha$                & Fixed                          & $2$                \\
$U_{\rm min}$           & $\mathcal{U}(0.1, 25)$         & $[0.1, 10]$         \\
$U_{\rm max}$           & Fixed                          & $3 \times 10^5$    \\
$\gamma$                & $\mathcal{U}(0, 1)$            & $[0, 0.5]$         \\
$q_{\rm PAH}$           & Fixed                          & $0.0047$           \\
$\tau_{9.7}$            & Fixed                          & $7$                \\
$\cos i_{\rm AGN}$      & $\mathcal{U}(0.75, 1)$             & $[0.75, 1]$            \\
\cutinhead{AGN Model $-$ No X-ray Component}
$\log L_{AGN}$          & $\mathcal{U}(11, 13)$           & $[12.0, 12.5]$    \\
\cutinhead{AGN Model $-$ With X-ray Component}
$N_{\rm H}$         & $\mathcal{U}(10^{-4}, 10^{5})$ & $[10^0, 10^2]$     \\          
$M_{\rm SMBH}$          & $\mathcal{U}(10^5, 10^{10})$   & $[10^6, 10^8]$     \\
$\log \dot m$           & $\mathcal{U}(-1.5, 0.3)$       & $[-1.0, 0.0]$      \\
\enddata
\tablecomments{$\mathcal{U}(a, b)$ indicates a uniform distribution from $a$ to $b$. Fixed parameters have their value listed in the
initialization range column.}
\end{deluxetable}

%% file: table_ngc4631.tex
\begin{deluxetable}{lcc}[t]
\tabletypesize{\footnotesize}
\tablecolumns{3}
\tablecaption{\label{tab:NGC4631Param}Summary of Parameters Used in the NGC~4631 Example.}
\tablehead{\colhead{Parameter} & \colhead{Prior Function} & \colhead{Initialization Range}}
\startdata
\cutinhead{All Models}
$\psi_j$               & $\mathcal{U}(0, 10^3)$        & $[0, 10]$           \\
$\alpha$               & Fixed                         & $2$                 \\
$U_{\rm min}$          & $\mathcal{U}(0.1, 25)$        & $[0.1, 5]$          \\
$U_{\rm max}$          & Fixed                         & $3 \times 10^5$     \\
$\gamma$               & $\mathcal{U}(0, 1)$           & $[0, 0.1]$          \\
$q_{\rm PAH}$          & $\mathcal{U}(0.0047, 0.0458)$ & $[0.0047, 0.0458]$  \\
\cutinhead{\citet{2000ApJ...533..682C} Models}
$\tau_{{\rm DIFF}, V}$ & $\mathcal{U}(0, 10)$          & $[0, 1]$            \\
\cutinhead{Inclination-dependent Models}
$\cos i$               & Tabulated\tablenotemark{a}    & $[0, 0.2]$          \\
$\tau_B^f$             & $\mathcal{U}(0, 8)$           & $[0, 4]$            \\
$B/D$                  & Fixed                         & $0$                 \\
$F$                    & $\mathcal{U}(0, 0.61)$        & $[0, 0.61]$         \\
\cutinhead{X-ray Models}
$N_{\rm H}$                  & $\mathcal{U}(10^{-4}, 10^5)$  & $[10^{-1}, 10^{2}]$ \\
\enddata
\tablecomments{$\mathcal{U}(a, b)$ indicates a uniform distribution from $a$ to $b$. Fixed parameters have their value listed in the
initialization range column.}
\tablenotetext{a}{Generated using the Monte Carlo method described in Section~3 of \citet{2022ApJ...931...53D}.}
\end{deluxetable}

%% file: table_ngc628_components.tex
\begin{deluxetable*}{l|lllllll}[ht]
\tabletypesize{\footnotesize}
\tablecolumns{8}
\tablecaption{\label{tab:NGC628Codes}Summary of the Components Used when Fitting with Each Bayesian Sampling SED-fitting Code.}
\tablehead{\colhead{Code} & \colhead{SSP} & \colhead{IMF} & \colhead{Metal.} & \colhead{SFH} & \colhead{Dust att.} & \colhead{Dust em.} & \colhead{Sampler}} 
\startdata
\texttt{Lightning}  & \texttt{P\`EGASE} & K01  & $Z=0.02$ & Non-param. & \citet{2000ApJ...533..682C} & \citet{2007ApJ...657..810D} & Affine-in. MCMC \\
\texttt{Prospector} & \texttt{MILES+MIST}     & K01  & $Z=0.02$ & Non-param. & \citet{2000ApJ...533..682C} & \citet{2007ApJ...657..810D} & Affine-in. MCMC \\
\texttt{BAGPIPES}   & BC03              & KB02 & $Z=0.02$ & Non-param. & \citet{2000ApJ...533..682C} & \citet{2007ApJ...657..810D} & Nested Sampling \\
\enddata
\tablecomments{Column (1): SED fitting code. Column (2): Stellar population synthesis models (\texttt{MILES+MIST} = \texttt{MILES} spectral library \citep{2011AA...532A..95F} with \texttt{MIST} isochones \citep{2016ApJS..222....8D, 2016ApJ...823..102C}; BC03 \citep{2003MNRAS.344.1000B}). Column (3): Initial mass function (K01 
\citep{2001MNRAS.322..231K}; KB02 \citep{2002MNRAS.336.1188K}).
Column (4): Metallicity, set to the specified constant value for all ages of SFH. Column (5): SFH form, all codes 
used the same age bins of 0~-- 10~Myr, 10~-- 100~Myr, 0.1~-- 1~Gyr, 1~-- 5~Gyr, and 5~-- 13.4~Gyr. Column (6): Dust attenuation 
curve, all codes used the base \citet{2000ApJ...533..682C} curve extrapolated to the Lyman limit. Column (7): Dust 
emission model, all codes used the \citet{2007ApJ...657..810D} model with energy balance and fixed $\alpha=2$.
\texttt{Prospector} and \texttt{BAGPIPES} have $U_{\rm max}=10^6$, while \texttt{Lightning} has 
$U_{\rm max}=3 \times 10^5$ (these differences in $U_{\rm max}$ have minimal effects on the models). Column (8): Bayesian 
sampler, \texttt{Lightning} and \texttt{Prospector} use the same affine-invariant MCMC algorithm, while \texttt{BAGPIPES} uses the \texttt{MultiNest} nested sampling algorithm.}
\end{deluxetable*}

%% file: table_ngc628.tex
\begin{deluxetable*}{lcccc}[t]
\tabletypesize{\footnotesize}
\tablecolumns{5}
\tablecaption{\label{tab:NGC628Param}Summary of Parameters Used in the NGC~628 Example.}
\tablehead{\colhead{Parameter} & \colhead{Prior Function} & \colhead{\texttt{Lightning} Init.\tablenotemark{a}} & 
           \colhead{\texttt{Prospector} Init.\tablenotemark{b}} & \colhead{\texttt{BAGPIPES} Init.\tablenotemark{c}}}
\startdata
\cutinhead{All Models}
$\tau_{{\rm DIFF}, V}$\tablenotemark{d} & $\mathcal{U}(0, 10)$          & $[0, 3]$           & $[1, 0.5]$     & \nodata \\
$\alpha$                                & Fixed                         & $2$                & $2$            & $2$     \\
$U_{\rm min}$                           & $\mathcal{U}(0.1, 25)$        & $[0.1, 10]$        & $[5, 1]$       & \nodata \\
$U_{\rm max}$                           & Fixed                         & $3 \times 10^5$    & $10^6$         & $10^6$  \\
$\gamma$                                & $\mathcal{U}(0, 1)$           & $[0, 0.5]$         & $[0.1, 0.5]$   & \nodata \\
$q_{\rm PAH}$                           & $\mathcal{U}(0.0047, 0.0458)$ & $[0.0047, 0.0458]$ & $[0.02, 0.01]$ & \nodata \\
\cutinhead{\texttt{Lightning} SFH}
$\psi_j$                                & $\mathcal{U}(0, 10^3)$        & [0, 10]            & \nodata        & \nodata \\
\cutinhead{\texttt{Prospector} and \texttt{BAGPIPES} SFH}
$M_j$\tablenotemark{e}                  & $\mathcal{U}(0, 10^{12})$     & \nodata            & $[10^7, 10^7]$ & \nodata \\
\enddata
\tablecomments{$\mathcal{U}(a, b)$ indicates a uniform distribution from $a$ to $b$. Fixed parameters have their value listed in the
initialization range columns.}
\tablenotetext{a}{The initialization range specified for \texttt{Lightning}.}
\tablenotetext{b}{The initialization parameters specified for \texttt{Prospector}. The first value is the median starting point, and
the second is the dispersion scale around that point.}
\tablenotetext{c}{The nested sampling algorithm in \texttt{BAGPIPES} does not require initialization. However, we list this
column to show the fixed parameter values.}
\tablenotetext{d}{\texttt{BAGPIPES} normalizes the \citet{2000ApJ...533..682C} curve with $A_V$, which we convert 
to $\tau_V$ via $\tau_V = 0.4 \ln(10) A_V$.}
\tablenotetext{e}{The stellar mass parameter in $M_\odot$. \texttt{Prospector} and \texttt{BAGPIPES} normalize their SFH 
bins to unit stellar mass, where \texttt{Lightning} normalizes to unit SFR. Therefore, the priors on the parameters are 
different.}
\end{deluxetable*}

%% file: table_CIGALEparams.tex
\begin{deluxetable*}{lcc}[t!]
\tabletypesize{\footnotesize}
\tablecolumns{3}
\tablecaption{\label{tab:CIGALE}\texttt{CIGALE} Gridded or Nondefault Parameters.}
\tablehead{\colhead{Module} & \colhead{Parameter} & \colhead{Values}}
\startdata
SFH; ${\rm SFR} \propto t \exp(-t/\tau)$       & Stellar e-folding time  & 0.1, 0.5, 1, 5 Gyr \\
                                               & Stellar age             & 0.5, 1, 3, 5,7 Gyr \\
\cline{1-3}
SSP; \citep{2003MNRAS.344.1000B}               & IMF                     & \citet{2003PASP..115..763C} \\
                                               & Metallicity, $Z$        & 0.02 \\
\cline{1-3}
Dust att.; \citet{2000ApJ...533..682C}  & Color excess, $E(B-V)$  & 0.05, 0.1, 0.2, 0.3, 0.4, 0.5, 0.7, 0.9 mag \\ 
\cline{1-3}
Dust em.; \citet{2014ApJ...784...83D}     & $\alpha$ slope in $dM_{\rm dust} \propto U^{-\alpha}dU$ & 1.5, 2.0, 2.5 \\
\cline{1-3}
AGN (UV-to-IR) SKIRTOR                         & AGN contribution to IR luminosity, ${\rm frac_{\rm AGN}}$ & 0.01 to 0.99 (step 0.1) \\
                                               & Wavelength range used to calculate $\rm frac_{AGN}$   & 5--1000 $\mu$m \\
                                               & Viewing angle, $\theta$                     & 0, 10, 20, 30, 40$^\circ$  \\
                                               & Polar-dust color excess, $E(B-V)$ & 0, 0.05, 0.1, 0.15, 0.2, 0.3 mag \\
                                               & Modified optical slope power-law index, $\delta_{\rm AGN}$  & $-1$ to 1 (step 0.25) \\
\cline{1-3}
X-ray                                          & Power-law slope, $\alpha_{\rm OX}$, of the $\tilde{L}_{2\rm\ keV}-\tilde{L}_{2500}$ relationship  & $-1.9$ to $-1.1$ (step 0.1) \\
\enddata
\tablecomments{For all other parameters not listed, their default values were utilized.}
\end{deluxetable*}

%% file: table_Lightnging_v_CIGALE.tex
\begin{deluxetable*}{l|ccccc}
\tablecaption{\label{tab:LightvCIGALE}Derived Parameters from \texttt{Lightning} and \texttt{CIGALE}.}
\tablecolumns{6}
\tablehead{Code & SFR                   & $\log_{10}$($M_\star$)   & $\log_{10}$($L_{\rm TIR}$) & $i_{\rm AGN}$ & $\log_{10}$($L_{\rm AGN}$) \\                & [$M_\odot$ yr$^{-1}$] & [$M_\odot$]              & [$L_\odot$]                & [degrees]     & [$L_\odot$]                \\           (1)  & (2)                   & (3)                      & (4)                        & (5)           & (6)                 }
\startdata
\texttt{Lightning} & $47.7^{+34.6}_{-20.9}$ & $10.64^{+0.22}_{-0.24}$ & $11.98^{+0.05}_{-0.05}$ & $40.3^{+0.9}_{-6.6}$ & $12.10^{+0.04}_{-0.04}$ \\
\texttt{CIGALE}    & $68.6 \pm 17.4$ & $10.70 \pm 0.43$ & $11.90 \pm 0.10$ & $33.1 \pm 10.7$ & $12.26 \pm 0.07$ \\
\tableline
Code & ${\rm frac}_{\rm AGN}$ & $\alpha_{\rm OX}$ & $\log_{10}$($M_{\rm SMBH}$) & $\log_{10} \dot m$ & $\delta_{\rm AGN}$ \\     &                        &                   & [$M_\odot$]                 &                    & \\ (1) & (7)                    & (8)               & (9)                         & (10)               & (11) \\
\tableline
\texttt{Lightning} & $0.45^{+0.03}_{-0.04}$ & $-1.29^{+0.02}_{-0.01}$ & $9.05^{+0.02}_{-0.02}$ & $-1.45^{+0.02}_{-0.02}$ & \nodata \\
\texttt{CIGALE}    & $0.47 \pm 0.13$ & $-1.32 \pm 0.04$ & \nodata & \nodata & $-0.18 \pm 0.26$ \\
\enddata
\tablecomments{For \texttt{Lightning}, the uncertainties are listed as the 16th and 84th percentile range. For \texttt{CIGALE}, the uncertainties are listed as the derived 1$\sigma$ uncertainties. Column (1): SED fitting code. Column (2): Recent SFR of the last 100 Myr. Column (3): Surviving stellar mass. Column (4): Total integrated IR luminosity of the dust emission model. Column (5): Line-of-sight AGN torus inclination. Column (6): UV-to-IR AGN bolometric luminosity. Column (7): Fraction of the AGN IR luminosity to the total IR luminosity between observed-frame 5--1000~$\mu$m. Column (8): Slope of the $\tilde{L}_{2\rm\ keV}-\tilde{L}_{2500}$ relationship ($\alpha_{\rm OX} = -0.3838 \log_{10}(L_{2500{\rm \AA}}/L_{2{\rm keV}})$). Column (9): SMBH mass (unique to \texttt{Lightning}). Column (10): SMBH accretion rate, Eddington rate normalized (unique to \texttt{Lightning}). Column (11): Modified optical slope power-law index of the SKIRTOR templates.}
\end{deluxetable*}